\documentclass[sigconf,nonacm]{acmart}
\setcopyright{none}
\AtBeginDocument{%
  }

\usepackage[dvipsnames]{xcolor}
\usepackage[utf8]{inputenc} %
\usepackage[T1]{fontenc}    %
\usepackage{hyperref}       %
\usepackage{url}            %
\usepackage{booktabs}       %
\usepackage{pifont}        %
\usepackage{placeins}
\usepackage{amsfonts}       %
\usepackage{amsmath}       %
\usepackage{nicefrac}       %
\usepackage{microtype}      %
\usepackage{xcolor}         %
\usepackage{xspace}
\usepackage{graphicx}
\usepackage{float}
\usepackage{cuted}
\usepackage{subcaption}
\usepackage[most]{tcolorbox}
\usepackage{listings}
\usepackage{multirow}
\usepackage{pgffor}
\usepackage{xparse}
\usepackage{array}
\usepackage{wrapfig}
\usepackage{cleveref}
\usepackage{algorithm}
\usepackage{algpseudocode}
\usepackage{cuted}
\usepackage[checkfootnote]{flushend}

\tcbuselibrary{listingsutf8}

\newcommand{\pname}{\texttt{MAPLE}\xspace}
\newcommand{\pnameone}{\texttt{MAPLE-1Q}\xspace}
\newcommand{\methodname}{\texttt{MAPLE-Synth}\xspace}
\newcommand{\benchname}{\texttt{MAPLE}\xspace}
\newcommand{\cmark}{\textcolor{blue}{\ding{51}}}
\newcommand{\xmark}{\textcolor{red}{\ding{55}}}

\newcommand{\openreview}{OpenReview\xspace}
\newcommand{\eg}{\hbox{\emph{e.g.,}}\xspace}

\newcommand{\datasetlink}{\href{https://huggingface.co/datasets/kai-02/MAPLE}{https://huggingface.co/datasets/kai-02/MAPLE}}
\newcommand{\codelink}{\href{https://github.com/Ggballs/MAPLE}{https://github.com/Ggballs/MAPLE}}

\usepackage[table,xcdraw]{xcolor}

\newtcolorbox{biggreenbox}[1][]{%
  colback=green!10!white,
  colframe=green!60!black,
  title=#1,
  sharp corners=south,
  breakable,
  enhanced,
  boxrule=0.8pt,
  left=5pt,
  right=5pt,
  top=5pt,
  bottom=5pt,
}

\usepackage{fontawesome5} %

\title{Can Retrievers Find the Same Paper from Different Aspects? \texorpdfstring{\newline}{ }A Multi-Aspect Full-Paper Scientific Retrieval Benchmark}

\begin{document}

\author{Yiyang Wei}
\affiliation{%
  \institution{Zhejiang University}
  \city{Hangzhou}
  \country{China}}
\email{marswei@zju.edu.cn}

\author{Fang Guo}
\affiliation{%
  \institution{Westlake University}
  \city{Hangzhou}
  \country{China}}
\email{guofang@westlake.edu.cn}

\author{Qiji Zhou}
\affiliation{%
  \institution{Westlake University}
  \city{Hangzhou}
  \country{China}}
\email{zhouqiji@westlake.edu.cn}

\author{Zhizhang Fu}
\affiliation{%
  \institution{Westlake University}
  \city{Hangzhou}
  \country{China}}
\email{fuzhizhang@westlake.edu.cn}

\author{Mengru Ding}
\affiliation{%
  \institution{Alibaba Group}
  \city{Beijing}
  \country{China}}
\email{dingmengru.dmr@alibaba-inc.com}

\author{Kai Yang}
\affiliation{%
  \institution{Shanghai Jiao Tong University}
  \city{Shanghai}
  \country{China}}
\email{erpound@sjtu.edu.cn}

\author{Yue Zhang}
\authornote{Corresponding author.}
\affiliation{%
  \institution{Westlake University}
  \city{Hangzhou}
  \country{China}}
\email{zhangyue@westlake.edu.cn}




\begin{abstract}
Scientific papers contain multiple searchable facets such as background, methods. However, many paper retrieval benchmarks merely evaluate individual query-paper relevance, while overlooking other facets of the same paper.
To bridge this gap, we introduce \pname, an expert-validated benchmark for multi-aspect, full-paper retrieval that evaluates whether retrievers can consistently recover the same paper from queries targeting its motivation, method, and experimental findings. \pname contains 2,095 queries about recent ML and NLP papers, grounded in both textual and multimodal content. We further propose \methodname, a retrieval-based in-context learning pipeline that leverages \openreview discussions and human-written query exemplars to generate realistic queries reflecting researchers' interests in different aspects of a paper. 
Our expert validation shows that these queries are comparable in realism to human-written queries and highly relevant to the target papers. Experiments across lexical, scientific-domain, general-purpose text, and multimodal retrievers reveal a substantial gap between retrieving a paper from any one aspect and retrieving it from all aspects: the strongest model achieves 98.1\% AnyAspect@20 but only 15.7\% AllAspect@20. Experiment/result queries and table-referenced queries are particularly difficult across retrievers. Although multi-chunk aggregation improves multi-aspect paper retrieval, considerable failures persist. \pname provides a testbed for evaluating and developing retrievers that represent scientific papers more comprehensively.\footnote{Our dataset is available at \datasetlink}
\end{abstract}
\maketitle
\pagestyle{plain}


\begin{figure}[t]
    \centering
    \includegraphics[width=\columnwidth]{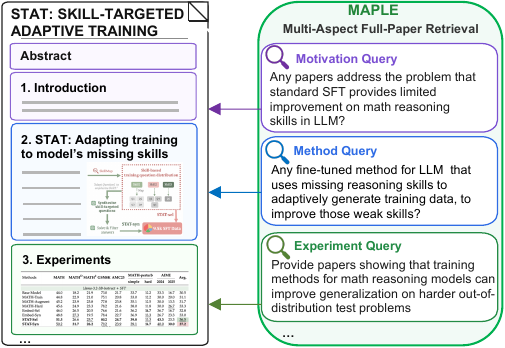}
    \caption{Example target paper and representative queries from \pname. The figure shows one query from each aspect group, including motivation, method, and experiment/result, which are grounded in different parts of the full paper, including textual and multimodal evidence. \pname evaluates whether a retriever can consistently recover the same paper across all  associated queries.}
    \label{fig:overview}
    \Description{}
\end{figure}

\section{Introduction}
\label{sec:intro}

\begin{table*}[!t]
    \centering
    \scriptsize
    \resizebox{\textwidth}{!}{\begin{tabular}{l|ccccc|>{\centering\arraybackslash}p{0.48in}|>{\centering\arraybackslash}p{0.40in}@{\hspace{1pt}}>{\centering\arraybackslash}p{0.40in}}
    \toprule
    \textbf{Benchmark} & \textbf{\#Query} & \shortstack{\textbf{Query}\\\textbf{Type}} & \shortstack{\textbf{Query}\\\textbf{Source}} & \shortstack{\textbf{Num \&}\\\textbf{Candidate Unit}} & \textbf{Modality} & \shortstack{\textbf{Many Query}\\\textbf{to One Target?}} & \multicolumn{2}{c}{\begin{tabular}{@{}>{\centering\arraybackslash}p{0.40in}@{\hspace{1pt}}>{\centering\arraybackslash}p{0.40in}@{}}\multicolumn{2}{@{}c@{}}{\textbf{Questions Based on}}\\Figs \& Tabs & Full-Text\end{tabular}} \\
    \midrule
    SciRepEval-SRCH~\cite{singh-etal-2023-scirepeval} & 530K & IR & Search Log & 5.31M abstracts & Text & \xmark & \xmark & \xmark \\
    DORIS-MAE~\cite{wang2023scientificdoris} & 100 & QA & Papers & 363K abstracts & Text & \xmark & \xmark & \xmark \\
    LitSearch~\cite{ajith-etal-2024-litsearch} & 597 & IR & Papers & 64K papers & Text & \xmark & \xmark & \cmark \\
    PaSa-RealScholarQuery~\cite{he-etal-2025-pasa} & 50 & IR & Search Log & $\sim$3.8K abstracts & Text & \xmark & \xmark & \xmark \\
    ArXivDoc~\cite{Khalighinejad2026DocumentasImageRF} & 547 & QA & Papers & 8.2K papers & Text + Image & \xmark & \cmark & \cmark \\
    IRPAPERS~\cite{Shorten2026IRPAPERSAV} & 180 & QA & Papers & 3.2K paper pages & Image & \xmark & \xmark & \xmark \\
    SciMMIR~\cite{wu-etal-2024-scimmir} & 530K & IR & Papers & 531K figures & Image & \xmark & \cmark & \xmark \\
    \rowcolor{red!5}
    \textbf{\pname (Ours)} & \textbf{2095} & \textbf{IR} & \textbf{OpenReview} & \textbf{74K papers} & \textbf{Text + Image} & \cmark & \cmark & \cmark \\
    \bottomrule
\end{tabular}}
    \caption{Comparison of scientific document retrieval benchmarks.}
    \label{tab:scientific_document_retrieval_benchmarks}
\end{table*}
Scientific paper retrieval takes a natural-language query describing a researcher's information need and returns a ranked list of papers that satisfy the query~\cite{ajith-etal-2024-litsearch,he-etal-2025-pasa}. 
Relevant benchmarks provide essential testbeds for evaluating and developing retrievers across abstract-level~\cite{singh-etal-2023-scirepeval}, full-text~\cite{ajith-etal-2024-litsearch}, and multimodal scientific retrieval~\cite{Shorten2026IRPAPERSAV,Khalighinejad2026DocumentasImageRF}. Recent retrievers achieve strong performance on many of these tasks~\cite{singh-etal-2023-scirepeval,muennighoff2025generative,jiang2025vlmvec}. For example, GritLM-7B achieves 80.0\% Recall@20 on LitSearch, improving to 85.2\% after reranking~\cite{zhang-etal-2025-scientific}. 
Despite recent progress in scientific paper retrieval, current retrievers still struggle to satisfy diverse scientific search needs in realistic settings~\cite{he-etal-2025-pasa}. One possible reason is that existing benchmarks still rely on a single-query-per-paper evaluation setting, evaluating each paper from only one scientific search intent~\cite{ajith-etal-2024-litsearch}. In reality, researchers often search for different aspects of the same paper, such as its methodology, experimental findings, implementation details, or downstream applications. For example, as illustrated in \Cref{fig:overview}, a retriever successfully retrieves a paper using a method-oriented query describing its adaptive training strategy, but fails when the query asks for experimental evidence of its generalization to unseen tasks. Consequently, strong benchmark performance under the conventional single-query-per-paper evaluation setting does not necessarily translate into reliable retrieval of the same paper across diverse scientific search intents.

Constructing a benchmark that reflects such realistic scientific search intents at scale is also challenging. Directly prompting an LLM with a target paper often yields artificial search intents or copies rare terminology from the paper itself~\cite{Shorten2026IRPAPERSAV,Khalighinejad2026DocumentasImageRF}, producing queries that do not reflect how researchers naturally search.

To evaluate whether retrievers can reliably retrieve the same paper across diverse scientific search intents, we introduce \textbf{\pname}, a benchmark measuring retrievers in
\textbf{\underline{M}}ulti-\textbf{\underline{A}}spect 
\textbf{\underline{P}}aper-centric 
fu\textbf{\underline{L}}l-text 
sci\textbf{\underline{E}}ntific retrieval. \pname contains 2,095 fine-grained queries associated with 210 recent machine learning papers and a retrieval corpus of 73,973 candidate papers. As illustrated in \Cref{fig:overview}, each target paper is paired with multiple queries grounded in textual and multimodal evidence and targeting distinct aspects of its motivation, method, and experimental findings. This many-query-to-one-paper design evaluates whether retrievers can consistently recover the same paper across its associated aspect queries.

To construct such a benchmark at scale with realistic scientific search intents, we introduce \textbf{\methodname}, a human-informed pipeline for scalable and realistic dataset generation. \methodname uses \openreview discussions as a source of reviewer-salient evidence and summarizes them into grounded aspect bullets. It then retrieves human-written, fine-grained search queries as in-context exemplars to guide LLM-based query generation.
Therefore, the generated queries preserve the scientific points that researchers consider important, while better matching natural human search language.
Expert validation shows that 96.7\% of sampled query--paper pairs are
relevant and that generated queries are comparable in realism to
human-written scientific search queries.


Our experiments reveal cross-aspect retrieval failures that are not captured by existing scientific retrieval benchmarks. 
Across lexical, scientific-domain, general-purpose textual and multimodal retrievers, 
models often retrieve the target paper for at least one aspect query while failing to retrieve it consistently across all associated aspects. For example, GritLM-7B~\cite{muennighoff2025generative} achieves 98.1\% AnyAspect@20 but only 15.7\% AllAspect@20.
This result is consistent with the fact that current retrievers achieve high performance on existing benchmarks but still cannot satisfy the full range of realistic information needs.
Our analyses further show that queries grounded in experimental evidence or tables are particularly challenging, especially when they provide limited descriptive constraints. To capture this fine-grained evidence in full-paper content, multi-chunk aggregation can improve performance, although
substantial failures remain. Finally, model performance on \pname is considerably lower than on established general and scientific retrieval benchmarks~\cite{ajith-etal-2024-litsearch,Khalighinejad2026DocumentasImageRF, muennighoff-etal-2023-mteb,thakur2021beir}, indicating that \pname provides a challenging testbed for developing more comprehensive scientific retrievers.

Our main contributions are summarized below:

\begin{enumerate}
    \item We introduce \benchname, an expert-validated benchmark for multi-aspect full-paper scientific retrieval that evaluates whether retrievers can consistently recover the same target paper from multiple fine-grained aspect queries.
    
    \item We propose \methodname, a retrieval-based in-context learning (ICL) framework that combines human-written exemplars with \openreview comments to generate realistic scientific search queries automatically.

    \item We conduct an evaluation that disentangles retrievers' comprehensive understanding of full-paper retrieval, and analyze performance across query aspects, evidence sources,  constraint count, and paper representation strategies.
\end{enumerate}

\section{Related Work}
\label{sec:related_work}
\noindent\textbf{Scientific Paper Retrieval Benchmarks}
\label{subsec:benchmark_scientific_retrieval_related_work}
Scientific paper retrieval benchmarks have evolved from domain-specific ad-hoc retrieval such as TREC Genomics~\cite{DBLP:conf/trec/HershCRR06} and TREC-COVID~\cite{10.1145/3451964.3451965} toward more realistic scientific evidence~\cite{wadden-etal-2020-fact, wadden-etal-2022-scifact} and complex query search settings~\cite{wang2023scientificdoris}.  SciRepEval~\cite{singh-etal-2023-scirepeval} summarizes prior IR datasets and introduces a multi-task scientific IR benchmark for comprehensive paper representation tasks. As coming into AI era, PaSa~\cite{he-etal-2025-pasa} brings this setting into the agentic search domain. However, these benchmarks focus on abstract-level search, ignoring the details in full-text paper content. To address this gap, LitSearch~\cite{ajith-etal-2024-litsearch} is proposed to evaluate the paper retrieval settings linking the queries to paper full-text content. When it comes to multimodal paper retrieval settings, recent multimodal benchmarks study scientific figures~\cite{wu-etal-2024-scimmir}, tables, PDF pages~\cite{Shorten2026IRPAPERSAV}, page images, or structured full-paper evidence~\cite{Khalighinejad2026DocumentasImageRF}. Beyond semantic relevance evaluation, some reasoning-intensive retrieval benchmarks~\cite{zhang2026mrmr, su2025bright} link challenging queries to scientific snippets that barely share semantic overlap and require reasoning to judge relevance.
Despite these advances, most existing benchmarks still evaluate each target paper using only a single query. Consequently, they cannot assess whether a retriever consistently retrieves the same paper across multiple scientific search intents (detailed comparison is shown in  \Cref{tab:scientific_document_retrieval_benchmarks}).
\noindent\textbf{Scientific Document Retrieval Models.}
\label{subsec:scientific_retrieval_model_related_work}
Scientific paper retrieval has evolved from lexical matching to increasingly expressive paper representations. Early approaches relied on lexical matching~\cite{robertson2009probabilistic}, while scientific-domain encoders adapted pretrained language models to scientific corpora~\cite{beltagy-etal-2019-scibert}. Citation-aware models such as SPECTER~\cite{cohan-etal-2020-specter} and SciNCL~\cite{ostendorff-etal-2022-neighborhood} further leveraged citation graphs to learn paper representations. Recent retrieval models have substantially expanded retrieval capabilities. SPECTER2~\cite{singh-etal-2023-scirepeval} introduces task-specific adapters for diverse retrieval tasks, while general-purpose embedding models such as GritLM~\cite{muennighoff2025generative} have achieved strong performance on scientific retrieval benchmarks~\cite{ajith-etal-2024-litsearch}. Multimodal retrievers further extend retrieval to scientific figures, page images, and interleaved documents~\cite{jiang2025vlmvec,lin2025mmembed,zhang2026mrmr,Khalighinejad2026DocumentasImageRF}. However, existing evaluations primarily measure retrieval performance under the conventional single-query-per-paper setting. As a result, it remains unclear whether these retrievers can consistently retrieve the same paper across diverse scientific search intents.

\begin{table*}[t]
  \centering
  \scriptsize
  \begin{tabular}{p{0.06\textwidth}p{0.06\textwidth}p{0.35\textwidth}p{0.20\textwidth}p{0.22\textwidth}}
    \toprule
    \textbf{Subset} & \textbf{Aspect} & \textbf{\openreview Comment} & \textbf{Summarized Bullet} & \textbf{Generated Query} \\
    \midrule
    \parbox[c]{0.06\textwidth}{\centering Text-\\referenced}
    & Method
    & Review: ``... the policy is updated using \textcolor{BrickRed}{preference-based feedback} which takes the form of a \textcolor{BrickRed}{binary score between pairs of presented trajectories} ...'' \newline Comment: ``... offline imitation learning followed by \textcolor{BrickRed}{online preference-based fine-tuning} ... BRIDGE degrades gracefully with noisy feedback and maintains lower regret than the baseline ...''
    & The online phase uses \textcolor{BrickRed}{binary preference feedback} between presented \textcolor{BrickRed}{trajectories} to guide policy improvement.
    & Are there papers that learn control policies from \textcolor{BrickRed}{pairwise human preferences} over \textcolor{BrickRed}{trajectories}? \\
    \midrule
    \parbox[c]{0.06\textwidth}{\centering Multimodal-\\referenced}
    & \parbox[c]{0.06\textwidth}{\centering Experiment\\/Result}
    & Review: ``... The performance gains, especially for \textcolor{BrickRed}{2-bit quantization, are dramatic} (\textbf{\textcolor{blue}{Table 2}}). The method \textcolor{BrickRed}{successfully retains performance} where \textcolor{BrickRed}{SOTA PTQ methods} (GPTQ, AWQ) \textcolor{BrickRed}{fail completely}. ...''
    & \textbf{\textcolor{blue}{Table 2}} shows \textcolor{BrickRed}{dramatic 2-bit empirical gains over PTQ baselines}, including GPTQ and AWQ.
    & Could you list research showing that \textcolor{BrickRed}{2-bit quantization} can preserve reasoning accuracy \textcolor{BrickRed}{better than common post-training quantization methods} for language models? \\
    \bottomrule
\end{tabular}
  \caption{Representative text- and multimodal-referenced query-generation examples. Colored text highlights the alignment among generated queries, summarized bullets, and supporting \openreview comments. Please refer to \Cref{tab:query_generation_examples} for a complete illustration covering all query aspect groups and evidence types.}
  \label{tab:query_generation_examples_main}
\end{table*}

\section{\pname Dataset}
\label{sec:dataset}
\begin{figure}[t]
    \centering
    \includegraphics[width=\columnwidth]{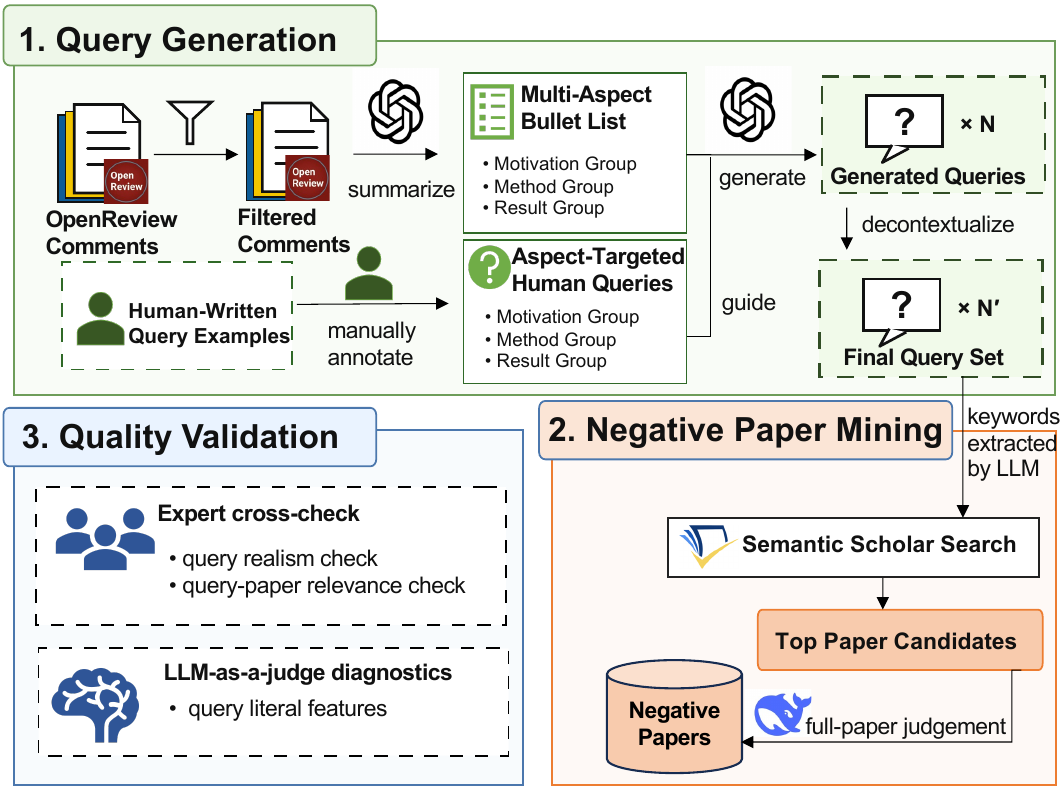}
    \caption{Overview of the \pname data construction pipeline, including query generation, negative mining and quality validation stages.}
    \label{fig:data_collecation}
    \Description{}
\end{figure}

Our benchmark \pname consists of (a) a large corpus of scientific papers $P$ including positive target papers $P_+$ and negatives $P_-$, and (b) synthetic paper search queries $Q$ targeting each  $P_+$. Each query in \pname represents one fine-grained aspect of its
target paper, and is assigned to one of three aspect groups: motivation, method, or experiment/result.
In this section, we describe the construction of \pname (see \Cref{fig:data_collecation}), including query generation (\Cref{subsec:query_gen}), negative paper mining (\Cref{subsec:neg_mining}), and quality validation (\Cref{subsec:quality_valid}). We construct two complementary splits: a main text-referenced split for full-paper multi-aspect retrieval, and a multimodal-referenced subset derived from comments that explicitly reference multimodal evidence (\eg figures, tables). Data statistics are reported in \Cref{tab:data_construction_statistics}.

\subsection{Query Generation}
\label{subsec:query_gen}
The goal of query generation is to create realistic scientific paper-search queries that (a) resemble human-written search-box inputs and (b) target non-trivial full-paper content. To achieve this goal, \methodname consists of an offline preparation step to construct a human-written query exemplar pool, followed by three automated stages, including (1) reviewer-salient bullet summarization, (2) retrieval-based ICL query synthesis, and (3) query decontextualization. Representative examples are shown in \Cref{tab:query_generation_examples_main}.

\subsubsection{Offline Preparation: Human-Written Query Pool Construction}
\label{subsubsec:human_query}
Before generating queries for target papers, we construct a pool of human-written scientific search queries from existing literature-search resources. Specifically, we collect high-quality queries from LitSearch~\cite{ajith-etal-2024-litsearch} and real scholar queries from PaSa~\cite{he-etal-2025-pasa}.
Since human paper-search queries vary in granularity, we manually filter this pool to retain fine-grained full-paper queries that express specific scientific information needs and exclude broad topic-level queries that only require abstract-level information~\cite{singh-etal-2023-scirepeval}. For each retained human query, we assign an aspect label according to the question: \textit{"What aspect of a paper would the retriever need to inspect to judge relevance?"} In terms of the labeled aspects, we follow \citet{baek-etal-2025-researchagent} and \citet{park2025chain}, a research paper could encapsulate multiple searchable facets, such as \textit{research motivation}, \textit{method design}, and \textit{experimental results}. The human-annotated distribution of queries across these three aspects is listed in \Cref{tab:section_aspect_label_criteria}. These labeled human queries serve as style exemplars, preparing for retrieval-based ICL during query generation. Detailed selection criteria and aspect labeling protocol are provided in Appendix~\ref{appendix:analysis_human_query}.

\subsubsection{Multi-Aspect Bullet Summarizing}
\label{subsubsec:multi-asp}
This stage summarizes detailed reviewer comments into concise literature-review-level descriptions, reducing the risk that downstream queries copy rare terminology or overly paper-specific phrasing.
We select \openreview discussions\footnote{https://openreview.net/} as the main source of reviewer-salient paper aspects. Unlike approaches that generate queries directly from paper text~\cite{ajith-etal-2024-litsearch, Shorten2026IRPAPERSAV, Khalighinejad2026DocumentasImageRF}, \openreview discussions highlight aspects that expert readers consider important, empirically meaningful, or discussion-worthy. This makes them useful for constructing detailed search intents that go beyond generic paper summaries. 
Specifically, we use ICLR 2026 as the generation source. Before summarization, we first apply a filtering stage to retain only accepted paper forums whose \openreview discussions explicitly reference multimodal
evidence, such as tables or figures. This filtering stage provides basic quality control and supports downstream multimodal-referenced query generation. 
For each retained paper forum, we collect its \openreview discussions and summarize them into aspect-specific bullet points. The summarization prompt (see \Cref{prompt:summarize_bullets}) is conditioned on the definitions of the three aspect groups: motivation, method, and experiment/result. Each bullet is required to be grounded in the discussion and to describe a scientific aspect that supports a realistic paper-search query.

\subsubsection{Retrieval-Based Query Synthesis}
\label{subsubsec:query_gen}
Directly using detailed aspect bullets to generate search queries, however, can produce overly specific or less human-like queries~\cite{Shorten2026IRPAPERSAV}. To generate more realistic and natural queries, we leverage a retrieval-based ICL framework that uses human-written scientific search queries as exemplars to guide generation.
Before retrieval, for each aspect bullet, we generate a short \textit{query seed} that captures the core search intent in natural language. The seed summarizes paper-specific wording while preserving the scientific hook of the bullet.
Given a query seed and its aspect label, \methodname retrieves top-$5$ human-written exemplars from the same aspect group. The retrieved exemplars are then inserted into the query generation prompt as in-context demonstrations. This retrieval-based ICL design encourages generated queries to follow the style, length, and granularity of similar human scientific search queries. To further improve realism at the dataset level, we sample the number of generated queries for each aspect according to the aspect distribution observed in the human-written query pool (see Appendix~\ref{appendix:analysis_human_query}), resulting in a benchmark whose search-intent distribution more closely matches real scientific search behavior (see \Cref{tab:data_construction_statistics}(a)).

In addition to the text-referenced queries, we extend the same pipeline to construct a multimodal-referenced subset. We identify \openreview discussions that explicitly reference figures, tables, equations, or other non-textual evidence, summarize them into multimodal aspect bullets using a specialized prompt (see \Cref{prompt:summarize_bullets_multimodal}), and generate queries through the same retrieval-based ICL framework.

\subsubsection{Query Decontextualization}
\label{subsubsec:query_decontext}
After query generation, we filter out queries that depend excessively on target-specific wording or whose relevance can be determined from the title and abstract alone. Specifically, we use a reasoning LLM~\cite{DeepSeekAI2026DeepSeekV4TH} to compare each generated query against the target paper's title and abstract using the rubric in \Cref{prompt:fullpaper_retrieval_evaluation}.
This decontextualization stage assigns a high-lexical-overlap tag to 36.1\% of the generated queries, which are subsequently removed. The remaining queries are more open-ended and require evidence from the full paper to establish relevance.

\subsection{Negative Paper Mining} 
\label{subsec:neg_mining}
For each query, the paired target paper is treated as the positive document. To construct a realistic candidate pool, we include two types of negative papers: (1) \textit{hard negatives}, which are topically similar papers but do not satisfy all detailed conditions in the query~\cite{su2025bright}, and (2) \textit{normal negatives}, which are sampled from a broad scientific corpus to approximate the background distribution in realistic paper retrieval scenario~\cite{ajith-etal-2024-litsearch}.

\subsubsection{Hard-Negative Mining}
Given a generated query, we first use a reasoning LLM~\cite{DeepSeekAI2026DeepSeekV4TH} to extract search keywords and semantic constraints. We submit these keywords to the Semantic Scholar Search bulk API\footnote{https://api.semanticscholar.org} and retain the top-$100$ retrieved candidate papers. We further leverage DeepSeek-V4 Pro~\cite{DeepSeekAI2026DeepSeekV4TH} to rerank the candidates and keep the top-30 papers for full-text validation.
To reduce false-negative risk, we parse each candidate paper with Docling~\cite{Auer2024DoclingTR} and provide its full text to a long-context reasoning LLM~\cite{DeepSeekAI2026DeepSeekV4TH}. The model judges whether the candidate paper satisfies the query. Candidates judged as positive are discarded, while candidates judged as topically similar but condition-mismatched are retained as hard negative papers. This process yields negative papers that are close enough to challenge retrieval systems while reducing the chance of labeling true positives as negatives.

\subsubsection{Normal-Negative Mining}
To simulate a realistic retrieval background, we additionally sample papers from ACL Anthology\footnote{https://aclanthology.org/} to be normal negative papers. Since the target papers cover multiple machine learning areas, especially language models and related subfields (see \Cref{tab:target_paper_primary_area}), ACL Anthology provides a broad but domain-relevant candidate pool. We sample more than 50K papers as normal negatives and combine them with the target papers and mined hard negatives to form the final retrieval corpus.

\subsection{Quality Validation}

\begin{table}[t]
  \centering
  \scriptsize
  \begin{subtable}[t]{\columnwidth}
    \centering
    {\begin{tabular}{p{0.2\linewidth}rrrr}
\toprule
\textcolor{black}{\textbf{Aspect Group}} & \textcolor{black}{\textbf{\#Q}} & \textcolor{black}{\textbf{\#MM (\%)}} & \textcolor{black}{\textbf{Q/$P_+$}} & \textcolor{black}{\textbf{Avg. Len.}} \\
\midrule
Motivation & 224 & 14 (6.3) & 1.1 & 23.5 \\
Method & 922 & 82 (8.9) & 4.4 & 24.0 \\
Experiment/Result & 949 & 319 (33.6) & 4.5 & 24.5 \\
\midrule
Total & 2095 & 415 (19.8) & 10.0 & 24.2 \\
\bottomrule
\end{tabular}

}
    \caption{Query statistics by aspect group. \#Q denotes the number of queries; \#MM (\%) denotes the number of multimodal-referenced queries, with the within-aspect percentage in parentheses; Q/$P_+$ denotes the average number of queries per positive paper. 
    Avg. Len. denotes the average query length in word count.}
    \label{tab:query_aspect_statistics}
  \end{subtable}

  \vspace{0.6em}
  \begin{subtable}[t]{\columnwidth}
    \centering
    {\setlength{\tabcolsep}{1pt}%
\begin{tabular}{p{0.20\linewidth}rrp{0.58\linewidth}}
\toprule
\textcolor{black}{\textbf{Paper Group}} & \textcolor{black}{\textbf{\#P}} & \textcolor{black}{\textbf{Avg. Len.}} & \textcolor{black}{\textbf{Source}} \\
\midrule
Positives & 210 & 15413.7 & ICLR 2026 papers \\
Hard negatives & 23739 & 10890.1 & ArXiv papers retrieved by Semantic Scholar search\\
Normal negatives & 50024 & 8710.0 & ACL Anthology papers \\
\midrule
Total & 73973 & 9428.7 & -- \\
\bottomrule
\end{tabular}}
    \caption{Paper statistics by label group. \#P denotes the number of papers; Avg. Len. denotes the average paper length in word count.}
    \label{tab:paper_label_statistics.}
  \end{subtable}
  \caption{Query and paper statistics for \pname construction.}
  \label{tab:data_construction_statistics}
\end{table}
\label{subsec:quality_valid}
After query generation and paper mining, the \pname construction process is complete, whose key statistics are listed in \Cref{tab:data_construction_statistics}. To validate the dataset quality, we conduct expert cross-checking for query realism and query--positive relevance (complete annotation protocol in Appendix~\ref{appendix:human_cross_check}). Additionally, we leverage an LLM-as-a-judge diagnostic system to characterize generated queries at scale.
\subsubsection{Expert Cross-Check}

\begin{table}[t]
  \centering
  \scriptsize
  {\setlength{\tabcolsep}{2pt}%
\begin{tabular}{p{0.38\linewidth}p{0.12\linewidth}p{0.12\linewidth}p{0.12\linewidth}p{0.16\linewidth}}
\toprule
\textcolor{black}{\textbf{Comparison}} & \textcolor{black}{\textbf{Prefer A}} & \textcolor{black}{\textbf{Tie}} & \textcolor{black}{\textbf{Prefer B}} & \textcolor{black}{\textbf{Cohen's $\kappa$}} \\
\midrule
\pname (ours) vs. Human & 12.0 & \textbf{78.0} & 10.0 & 0.24 \\
\midrule
\pname (ours) vs. LLM+Human Editing & 30.0 & \textbf{58.0} & 12.0 & 0.16 \\
Human vs. LLM+Human Editing & 26.0 & \textbf{60.0} & 14.0 & 0.19 \\
\midrule
\pname (ours) vs. LLM & \textbf{88.0} & 10.0 & 2.0 & 0.78 \\
Human vs. LLM & \textbf{82.0} & 8.0 & 10.0 & 0.61 \\
\bottomrule
\end{tabular}}
  \caption{Blinded pairwise query realism comparison results (values in \%). A and B denote the first and second query sources named in each comparison; Cohen's $\kappa$ reports inter-annotator agreement for each setting.}
  \label{tab:query_realism_main}
\end{table}
\paragraph{Query Realism Check.}
To evaluate whether \methodname produces realistic fine-grained search queries, we conduct a blinded pairwise human evaluation. We sample 150 generated queries and compare them against three query sources: (1) human-written queries, (2) LLM-generated queries that were subsequently reviewed or edited by humans, and (3) fully LLM-generated queries. We further include two calibration settings comparing human-written queries against the two LLM-based sources (\Cref{tab:query_realism_source_labels} summarizes all query sources). For each query pair, two annotators independently judge which query is more likely to be entered into a scientific paper search box, with disagreements resolved by a third annotator.

As shown in \Cref{tab:query_realism_main}, \methodname produces queries whose realism is comparable to human-written queries. When compared directly against human-written queries, 78.0\% of pairwise comparisons result in ties, while the remaining preferences are nearly balanced between the two sources. Similarly, comparisons among \methodname, human-written, and human-edited LLM queries are also dominated by ties. The corresponding Cohen's $\kappa$ values are modest (0.16--0.24), indicating that annotators often found these query sources difficult to distinguish and that differences in perceived realism were subtle.
In contrast, both \methodname and human-written queries are consistently preferred over fully LLM-generated queries, with substantially higher inter-annotator agreement ($\kappa=0.78$ and $0.61$, respectively). Together, these results demonstrate that \methodname generates search queries with realism comparable to human-written or human-edited queries, while substantially outperforming direct LLM generation.

\paragraph{Query-Paper Relevance.}
To verify that \methodname 's multi-stage generation pipeline preserves query--paper relevance despite its multiple generation steps, we manually evaluate query--paper relevance. Specifically, we sample 150 generated queries and ask experts to judge whether each query is supported by and relevant to its paired paper's full text. Overall, 96.7\% of the sampled query--paper pairs are judged relevant. This result indicates that the multi-stage generation pipeline generally preserves the intended relevance relationship, thereby supporting the reliability of the benchmark.

\begin{table*}[t]
    \centering
    \scriptsize
    \resizebox{0.82\textwidth}{!}
    {\setlength{\tabcolsep}{2.5pt}%
\begin{tabular}{lcccccccccccc}
\toprule
& \multicolumn{6}{c}{\textbf{Paper-level Metrics (\%)}} & \multicolumn{6}{c}{\textbf{Query-level Metrics (\%)}} \\
\cmidrule(lr){2-7}\cmidrule(lr){8-13}
\multirow{2}{*}{\textbf{Model}} & \multicolumn{2}{c}{\textbf{AllAspect}} & \multicolumn{2}{c}{\textbf{AnyAspect}} & \multicolumn{2}{c}{\textbf{AspectCoverage}} & \multicolumn{2}{c}{\textbf{Motivation}} & \multicolumn{2}{c}{\textbf{Method}} & \multicolumn{2}{c}{\textbf{Experiment/Result}} \\
\cmidrule(lr){2-3}\cmidrule(lr){4-5}\cmidrule(lr){6-7}\cmidrule(lr){8-9}\cmidrule(lr){10-11}\cmidrule(lr){12-13}
& @5 & @20 & @5 & @20 & @5 & @20 & R@5 & R@20 & R@5 & R@20 & R@5 & R@20 \\
\midrule
BM25 & 0.0 & 0.0 & 68.6 & 89.0 & 15.4 & 28.3 & 16.5 & 33.0 & 18.8 & 32.1 & 11.5 & 22.8 \\
SPECTER2 & 0.0 & 1.4 & 38.1 & 55.2 & 12.3 & 22.5 & 17.9 & 31.2 & 15.7 & 25.2 & 8.0 & 18.1 \\
SciNCL & 0.5 & 1.4 & 63.3 & 75.7 & 17.8 & 30.7 & 23.2 & 35.3 & 18.5 & 33.1 & 15.2 & 26.3 \\
BGE-M3 & 1.0 & 2.9 & 64.3 & 80.0 & 21.9 & 33.3 & 30.8 & 42.9 & 24.7 & 36.3 & 16.8 & 27.6 \\
Instructor-XL & 0.0 & 2.9 & 68.1 & 80.5 & 21.0 & 34.1 & 32.1 & 50.9 & 24.0 & 37.2 & 14.8 & 26.4 \\
Ops-MM-Embed-7B & 2.4 & 5.2 & 59.5 & 82.4 & 21.3 & 40.3 & 24.1 & 43.8 & 23.0 & 42.1 & 18.4 & 37.8 \\
Qwen3-Embed-8B & 3.3 & 11.0 & 75.7 & 91.0 & 29.9 & 53.3 & 33.0 & 61.2 & 33.0 & 57.5 & 24.6 & 45.8 \\
Qwen3-VL-Embed-8B & \textbf{5.2} & 11.4 & 81.0 & 91.4 & 37.2 & 57.3 & 37.9 & 59.8 & 41.1 & 61.5 & 32.7 & 52.7 \\
GritLM-7B & 3.3 & \textbf{15.7} & \textbf{89.5} & \textbf{98.1} & \textbf{43.0} & \textbf{61.8} & \textbf{61.2} & \textbf{79.0} & \textbf{45.6} & \textbf{64.6} & \textbf{35.3} & \textbf{54.5} \\
\bottomrule
\end{tabular}}
    \caption{Retrieval performance on \pname. Paper-level metrics evaluate whether systems retrieve target papers across all, any, or covered aspects, while query-level metrics report specific recall rate inside each aspect group, including motivation, method, and experiment/result. R@k represents Recall@k.}
    \label{tab:retrieval_results}
\end{table*}
\subsubsection{LLM-as-a-Judge Diagnostics}
\label{subsubsec:llm_diagnostic}
To investigate literal feature similarity between generated queries and human-written ones, we leverage a rubric-based LLM-as-a-judge approach~\cite{zheng2023judging} to characterize query specificity, lexical naturalism, and semantic constraint count. 
We define semantic constraint count as the number of distinct conditions explicitly expressed in a query; it serves as a complementary quantitative measure of query specificity.
As shown in \Cref{fig:llm_judge_comparison}, \pname and human-written queries exhibit similar distributions, with moderate specificity and natural wording. In contrast, purely LLM-generated queries contain more constraints and are shifted toward excessive specificity and overly formal wording. These results provide evidence that retrieval-based human exemplars make \pname queries more consistent with human scientific search behavior. The complete rubric, judging prompt, sampling procedure, and detailed distributions are provided in Appendix \ref{app:llm_judge}.

\section{Experiments}
\label{sec:experiments}
\subsection{Experiment Setup}
\label{subsec:experiment_setup}
We compare the performance of different retrieval systems on our \pname benchmark. We report recall-based metrics at different cutoffs, including paper-level metrics and query-level metrics:

\begin{enumerate}
    \item \textbf{Paper-level metrics:}
    Let $P$ be the set of target papers, $Q_p$ be the set of aspect queries for paper $p$, each query $q$ represents one distinct aspect
of the paper. And $\mathbf{1}[q \rightarrow_k p]$ indicate whether query $q$ retrieves paper $p$ in the top-$k$ results. We define the reported paper-level metrics as
    \begin{align*}
        \text{AllAspect@}k
        &= \frac{1}{|P|}\sum_{p \in P}
        \mathbf{1}\left[\forall q \in Q_p, q \rightarrow_k p\right], \\
        \text{AnyAspect@}k
        &= \frac{1}{|P|}\sum_{p \in P}
        \mathbf{1}\left[\exists q \in Q_p, q \rightarrow_k p\right], \\
        \text{AspectCoverage@}k
        &= \frac{1}{|P|}\sum_{p \in P}\frac{1}{|Q_p|}
        \sum_{q \in Q_p}\mathbf{1}\left[q \rightarrow_k p\right].
    \end{align*}
    \textbf{AllAspect@k} measures the proportion of target papers retrieved by all of their associated aspect queries. \textbf{AnyAspect@k} measures the proportion of target papers retrieved by at least one associated aspect query. \textbf{AspectCoverage@k} measures the average proportion of successful aspect queries per target paper. 
    These paper-level metrics measure whether a retriever can recover a target paper partially or completely across different aspect queries.
    

    \item \textbf{Query-level metrics:}
 These metrics evaluate retrieval performance for individual aspect queries. We report \textbf{query-level recall@k} within each aspect group, including motivation, method, and experiment/result groups, where a query is counted as successful if its target paper appears in the top-$k$ retrieved results. 
 These query-level metrics measure retrieval performance for individual queries across different aspect groups.
\end{enumerate}

\subsection{Retrieval Models}
\label{subsec:retrieval_baselines}
We evaluate four groups of retrievers: (1) \textbf{sparse lexical retriever:} BM25~\cite{robertson2009probabilistic}; (2) \textbf{scientific document retrievers:} SPECTER2~\cite{singh-etal-2023-scirepeval} and SciNCL~\cite{ostendorff-etal-2022-neighborhood}; (3) \textbf{general-purpose text retrievers:} BGE-M3~\cite{chen-etal-2024-m3}, Instructor-XL~\cite{su-etal-2023-one}, Qwen3-Embed-8B~\cite{Zhang2025Qwen3EA}, and GritLM-7B~\cite{muennighoff2025generative}; and (4) \textbf{multimodal retrievers:} Qwen3-VL-Embed-8B~\cite{Li2026Qwen3VLEmbeddingAQ} and Ops-MM-Embed-7B~\cite{lin2025mmembed}. The first three groups operate on textual document representations: we use extracted full-text content~\cite{Auer2024DoclingTR}  for representation and truncate it to each model's maximum input length. For multimodal retrievers, we represent each paper using its PDF page screenshots.
These settings follow each retriever's intended input modality and evaluate their practical capability for full-paper retrieval. Details are provided in Appendix \ref{appendix:retriever_detail}.

\subsection{Main Results}
\label{subsec:retrieval_results}
\textbf{Existing retrievers perform poorly on  multi-aspect full-paper retrieval.}  Results in \Cref{tab:retrieval_results} reveal that retrieving a target paper across all aspects is very challenging; even the strongest model, GritLM-7B, achieves only 15.7\% on AllAspect@20. Besides, there is a large gap between AllAspect@20 and AnyAspect@20, suggesting that current retrievers can often identify a relevant paper from at least one aspect, but they do not robustly recover the same paper when queries target different motivations, methods, and experimental evidence. This provides a signal that current retrievers could not comprehensively understand and represent every detail inside a holistic long scientific paper input.

\noindent\textbf{Aspect difficulty is consistent across retrievers.} Across most retrievers, motivation queries achieve the highest recall, followed by method and experiment/result queries. For example, GritLM-7B achieves 79.0\% Recall@20 on motivation queries, 64.6\% on method queries, and 54.5\% on experiment/result queries. Models such as Qwen3-Embed-8B and BGE-M3 follow the same pattern. Paper search queries express the information need in natural language; however, the supporting evidence for experiment/result queries is often encoded in numerical results or visual structures rather than stated explicitly in prose, which requires precise full-paper evidence localization and understanding. In contrast, motivation and method evidence typically appear in descriptive text earlier in the paper, making it easier for retrievers to identify.


\noindent\textbf{General-purpose text retrievers outperform scientific document retrievers.} The best-performing system is GritLM-7B, which outperforms SPECTER2 and SciNCL on both paper-level and query-level metrics. One likely reason is context length: our benchmark requires retrieving from full-text evidence beyond titles and abstracts, while scientific retrievers such as SPECTER2 and SciNCL are constrained to short inputs, typically preserving only title and abstract information. In contrast, the strongest general-purpose retrievers can encode much longer inputs, better covering full paper content. Another possible factor is instruction tuning, which helps models interpret detailed natural-language information needs, such as 
``Papers showing that reference-guided post-training improves downstream alignment performance for LLMs.''
These findings suggest that \pname evaluates not only scientific similarity, but also long-context and instruction-following retrieval ability.

\noindent\textbf{Multimodal retrievers do not dominate this task.} GritLM-7B, which encodes  full text remains the strongest retriever overall, and outperforms the screenshot-based multimodal models on most metrics. Although multimodal retrievers can directly access document layout and visual evidence, representing an entire paper as many page screenshots also introduces substantial irrelevant content that may dilute the relevant evidence. 
However, Qwen3-VL-Embed-8B outperforms its text-only counterpart, Qwen3-Embed-8B, on most metrics, suggesting the potential value of visual full-document representations. We further investigate this trade-off in \Cref{subsec:representation_analysis}.


\section{Analysis}
\label{sec:analysis}

We introduce the analysis from three levels: 
 \textit{(1) benchmark level}, we examine what
the multi-aspect evaluation reveals beyond existing scientific retrieval benchmarks and situate the benchmark within the broader IR landscape in \Cref{subsec:comparing_other_bench_analysis}. 
\textit{(2) query level}, we analyze the failure patterns of queries in partially successful papers across evidence sources and constraint counts in \Cref{subsec:failure_anlaysis}.  \textit{(3) paper level}, we study how different representations of papers affect performance in a multi-aspect full-paper retrieval setting in \Cref{subsec:representation_analysis}.

\subsection{Comparison with Existing IR Benchmarks}
\label{subsec:comparing_other_bench_analysis}
We first compare \pname with existing scientific retrieval benchmarks
to examine the value of its multi-aspect evaluation (\Cref{subsubsec:sci_ir_bench_compare}). We then compare our benchmark with general benchmarks to analyze its features in a broader context (\Cref{subsubsec:gen_ir_bench_compare}).

\subsubsection{Scientific IR Benchmark Comparison}
\label{subsubsec:sci_ir_bench_compare}
A scientific paper can contain multiple searchable aspects~\cite{zhang-etal-2022-multiview,do-etal-2025-multi}. However, existing scientific retrieval benchmarks often associate each target paper with only a single query--paper relevance relation, overlooking other potentially relevant aspects in the same paper. \pname addresses this limitation through a multi-aspect evaluation that associates each target paper with multiple aspect-specific queries. 

To test whether this design reveals failures hidden by single query--paper relevance evaluation, we construct \pnameone by randomly sampling one query per target paper for a controlled comparison. \pnameone and the full \pname benchmark share the same corpus, paper representations, retrievers, and ranking procedure. \pnameone follows the single query--paper relevance retrieval setting, whereas \pname AllAspect@k requires the target paper to be successfully retrieved by every associated query.

As shown in \Cref{tab:paper_search_bench_comparison}, model performance rankings are broadly consistent across LitSearch~\cite{ajith-etal-2024-litsearch}, DORIS-MAE~\cite{wang2023scientificdoris}, ArXivDoc~\cite{Khalighinejad2026DocumentasImageRF}, and our benchmarks.  Furthermore, performance on \pnameone is generally lower than on the existing benchmarks, indicating that even the single-query setting of \pname is challenging. More importantly, the giant gap between performance on \pnameone and the AllAspect score on \pname indicates that strong retrieval on individual queries does not ensure consistent retrieval performance across all aspects of a paper. Thus, \pname exposes cross-aspect
retrieval failures that conventional scientific retrieval benchmarks
do not directly measure. We further analyze failure patterns in \Cref{subsec:failure_anlaysis}

\begin{table}[t]
    \centering
    \scriptsize
    \begin{subtable}[t]{\linewidth}
        \centering
        \begingroup
\setlength{\tabcolsep}{4pt}%
\begin{tabular}{lcc|cc}
    \toprule
    Model
    & \shortstack{LitSearch\\R@20}
    & \shortstack{DORIS-MAE\\R@20}
    & \shortstack{MAPLE-1Q\\R@20}
    & \shortstack{MAPLE\\AllAspect@20} \\
    \midrule
    BM25
    & 39.90
    & 30.50
    & 28.10
    & 0.00 \\

    SPECTER2
    & --
    & 43.36
    & 22.86
    & 1.40 \\

    Instr.-XL
    & 56.50
    & --
    & 30.00
    & 2.90 \\

    GritLM-7B
    & 70.80
    & --
    & 60.00
    & 15.70 \\
    \bottomrule
\end{tabular}
\endgroup
        \caption{Text retrievers.}
        \label{tab:paper_search_bench_comparison_text}
    \end{subtable}

    \vspace{0.5em}
    \begin{subtable}[t]{\linewidth}
        \centering
        \begingroup
\setlength{\tabcolsep}{4pt}%
\begin{tabular}{lc|cc}
    \toprule
    Model
    & \shortstack{ArXivDoc\\nDCG@10}
    & \shortstack{MAPLE-1Q\\nDCG@10}
    & \shortstack{MAPLE\\AllAspect@10} \\
    \midrule
    Ops-MM-Embed-7B
    & 63.00
    & 19.40
    & 3.30 \\

    Qwen3-VL-Embed-8B
    & 69.00
    & 33.10
    & 8.60 \\
    \bottomrule
\end{tabular}
\endgroup

        \caption{Multimodal retrievers.}
        \label{tab:paper_search_bench_comparison_multimodal}
    \end{subtable}
    \caption{Comparison among existing paper search benchmarks, \pnameone and \pname. Results are from~\cite{ajith-etal-2024-litsearch, wang2023scientificdoris, Khalighinejad2026DocumentasImageRF}.}
    \label{tab:paper_search_bench_comparison}
\end{table}

\begin{table}[t]
    \centering
    \scriptsize
    \begin{tabular}{lccccc}
    \toprule
    \textbf{Model} & \textbf{MAPLE} & \textbf{MSMARCO} & \textbf{NQ} & \textbf{HotpotQA} & \textbf{SciFact}  \\
    \midrule
    BM25 & \textbf{12.46} & 22.80 & 32.90 & 60.30 & 66.50  \\
    Instructor-XL & \textbf{16.82} & 41.61 & 57.24 & 55.88 & 64.56  \\
    Qwen3-Embed-8B & \textbf{23.88} & 43.60 & 65.25 & 76.78 & 78.46 \\
    GritLM-7B & \textbf{35.82} & 41.96 & 70.30 & 79.40 & 79.17 \\
    \bottomrule
\end{tabular}
    \caption{Retrieval performance on general and scientific retrieval benchmarks. All scores are nDCG@10 $\times$ 100.}
    \label{tab:benchmark_comparison}
\end{table}

\begin{figure*}[t]
    \centering
    \includegraphics[width=\textwidth]{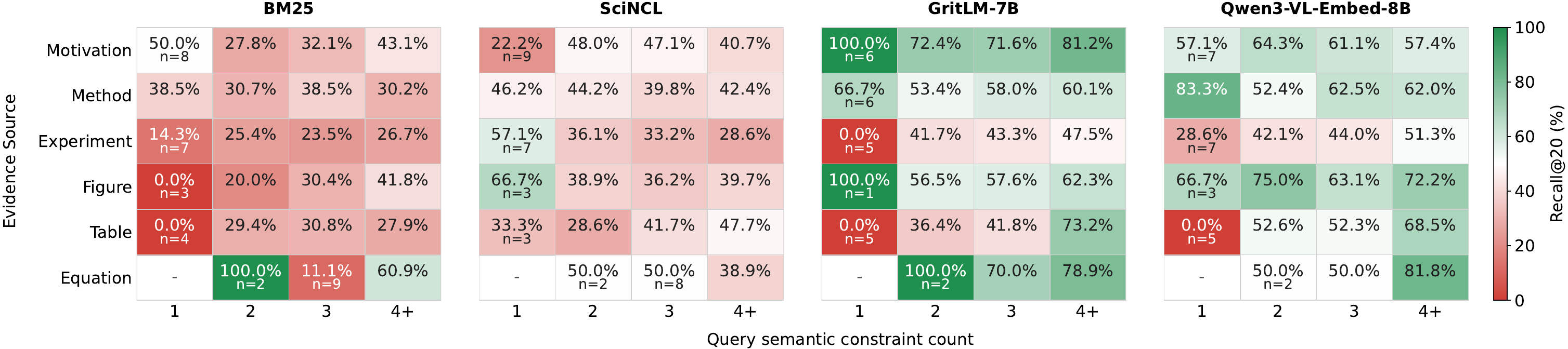}
    \caption{
Diagnostic analysis by evidence source and semantic-constraint for queries associated with partial-success papers. Each cell reports Recall@20, with colors ranging from dark red (0\%) to dark green (100\%). For groups containing fewer than 10 queries, the cell additionally reports the query count \(n\); these low-support estimates are interpreted cautiously in our analysis.
}
    \label{fig:partial_success_all_models_heatmap_grid}
\end{figure*}
\subsubsection{Comparing with General Retrieval Benchmarks}
\label{subsubsec:gen_ir_bench_compare}
Having established the diagnostic value of \pname within scientific paper retrieval, we next situate the benchmark within the broader IR landscape. 
We compare model performance on \pname to several popular general retrieval benchmarks, including MSMARCO~\cite{nguyen2017ms}, NQ~\cite{kwiatkowski-etal-2019-naturalNQ}, HotpotQA~\cite{yang-etal-2018-hotpotqa} and SciFact~\cite{wadden-etal-2020-fact}.
We collect publicly reported results from MTEB/BEIR-compatible sources~\cite{muennighoff-etal-2023-mteb, thakur2021beir} and official model reports. 
As shown in \Cref{tab:benchmark_comparison}, model scores on our \pname are substantially lower than those on other standard retrieval benchmarks, suggesting that \pname is considerably more challenging than general IR benchmarks.
Moreover, the model ranking on \pname follows the trend observed on more
complex retrieval tasks. Qwen3-8B and GritLM-7B are competitive on general
ad-hoc retrieval benchmarks: Qwen3-8B leads on MSMARCO, while GritLM-7B leads
on NQ. However, GritLM-7B is consistently stronger on more complex tasks such
as HotpotQA and SciFact. Our benchmark amplifies this gap, where GritLM-7B
outperforms Qwen3-8B by 10.00 nDCG@10 points. These results indicate that
\pname stresses deeper scientific relevance modeling beyond standard semantic matching.

\begin{figure*}[t]
    \centering
    \includegraphics[width=\textwidth]{\detokenize{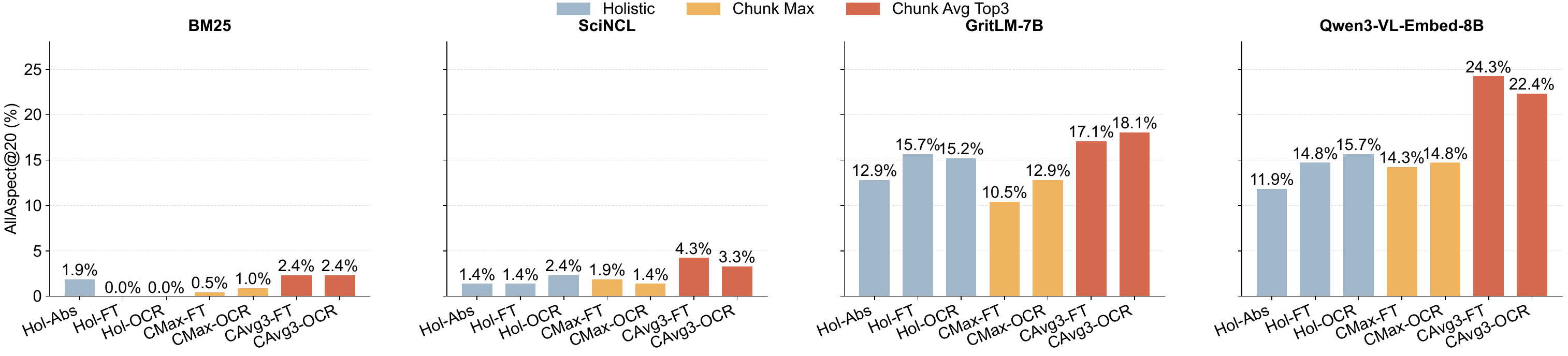}}
    \caption{AllAspect@20 performance (\%) under holistic and chunk-based paper representations  for BM25, SciNCL, GritLM-7B, and Qwen3-VL-Embed-8B. Hol-Abs, Hol-FT, and Hol-OCR denote holistic representations of the abstract, full text, and OCR-enhanced full text, respectively. CMax and CAvg3 denote max pooling and top-3 average pooling over document chunks.}
    \label{fig:representation_allaspect}
\end{figure*}

\subsection{Failure Analysis: What Causes Cross-Aspect Retrieval Failures?}
\label{subsec:failure_anlaysis}
Having established that multi-aspect evaluation reveals retrieval failures not captured by single-query-per-paper evaluation, we next investigate which query characteristics are associated with these failures. Although AnyAspect@20 indicates that a target paper can be retrieved from at least one query, 
failure under AllAspect@20 suggests that retrieval difficulty is different among aspect queries. 
We therefore analyze the \textit{partial-success} papers for each retriever, defined as \(\mathcal{P}_{\mathrm{partial}}^{20}(m)=\{p \mid \mathrm{AnyAspect@20}_{m}(p)=1 \land \mathrm{AllAspect@20}_{m}(p)=0\}\),
where \(m\) denotes the retrieval model. We select BM25, SciNCL, GritLM-7B, and Qwen3-VL-Embed-8B as representative lexical, scientific-domain, general-purpose text, and multimodal retrievers, respectively. For the queries associated with each model's partial-success papers, we report Recall@20 grouped by evidence source and semantic constraint count in \Cref{fig:partial_success_all_models_heatmap_grid}; the semantic constraint extraction process is described in \Cref{subsubsec:llm_diagnostic}. 

Contrary to the common intuition that more semantic constraints necessarily make retrieval harder, increasing the number of constraints does not consistently reduce retrieval performance. For GritLM-7B and Qwen3-VL-Embed-8B, recall often remains stable or even improves for sufficiently populated queries containing four or more semantic constraints, suggesting that additional constraints can provide useful semantic anchors rather than simply increasing query complexity. In contrast, BM25 and SciNCL exhibit clear performance degradation on highly constrained queries, particularly those referencing experimental evidence.

A much clearer difficulty pattern emerges when grouping queries by evidence source. Across all retrievers, motivation-referenced queries are generally easier, whereas experiment-referenced queries are consistently more difficult, indicating that the type of scientific evidence required by a query has a larger impact on retrieval success than its semantic complexity. Within multimodal evidence, table-referenced queries are consistently among the most challenging. This trend is particularly pronounced for embedding-based retrievers, where table-referenced queries containing only two semantic constraints remain substantially harder than many other query types despite having sufficient statistical support ($n \geq 10$). 

\subsection{Which Representation Best Captures Full-Paper Content?}
\label{subsec:representation_analysis}
In the main benchmark, each paper is encoded as a single holistic representation and results show substantial failure on AllAspect@20.
We next examine whether alternative paper representations can improve retrieval consistency across aspects. 
We compare holistic encodings of the abstract, full text, and OCR-enhanced full text~\cite{RapidOCR2021} with chunked encodings of the latter two formats. Motivated by prior evidence that plain fixed-size chunking is the most cost-effective approach~\cite{qu-etal-2025-semanticchunking}, we divide each paper into 512-token chunks and compare max pooling with top-$3$ average pooling.

As shown in \Cref{fig:representation_allaspect}, top-3 average pooling consistently outperforms the corresponding holistic representations across retrievers, whereas max pooling provides the weakest gains. Fixed-size chunks can reduce interference from unrelated paper content, but they may also fragment evidence that spans adjacent sections. Max pooling relies on a single highest-scoring chunk and can therefore miss distributed evidence. In contrast, top-3 average pooling aggregates several highly relevant chunks from different paper locations, improving evidence coverage while excluding irrelevant content. The effectiveness of paper representations, however, remains model-dependent: for BM25, the holistic abstract input outperforms holistic full-text and OCR-enhanced inputs, while embedding models generally benefit from richer full-paper representations. However, OCR provides only marginal and inconsistent gains over plain full text across retrievers. Extending the analysis to multimodal formats in \Cref{tab:representation_recall}, PDF screenshots with top-3 average pooling achieve the highest AllAspect@20 (24.76) and Experiment/Result Recall@20 (66.91). These results demonstrate the potential of visual page representations and multimodal retrievers for future tuning.

\section{Conclusion}
\label{sec:conclusion}
We propose \pname, a multi-aspect full-paper scientific retrieval benchmark that evaluates whether retrievers can consistently recover the same paper from multiple queries targeting its motivation, method, and experimental findings. We further introduce \methodname, a retrieval-based ICL pipeline that leverages human-written query exemplars and  \openreview discussions to generate realistic scientific search queries at scale. 
The expert validation shows that the generated queries
are comparably realistic to human queries and highly relevant to source papers. Experiments reveal substantial multi-aspect paper retrieval failures: even the strongest model achieves only 15.7\% AllAspect@20. Experiment/result and table-referenced queries are particularly challenging, indicating that current retrievers struggle to localize fine-grained empirical evidence across structural and numerical sources. Although multi-chunk aggregation improves full-paper representations, a considerable gap remains. We hope \pname provides a testbed for evaluating and advancing comprehensive, fine-grained full-paper retrieval.

\section*{Limitations and Future Work}
\label{sec:limitation}
Our data construction pipeline relies on publicly available \openreview discussions, and the current benchmark focuses on machine learning and computer science papers submitted to ICLR 2026. This scope may not fully capture the diversity and complexity of real-world paper search scenarios. Future work could extend \pname to biomedical and other scientific domains by incorporating domain-specific papers together with available peer-review reports or other expert-authored commentary. Moreover, because publicly available datasets of human-written full-paper search queries are limited, our exemplar pool remains relatively small. This limitation may reduce the diversity of the generated queries and constrain generalization across languages, search intents, and scientific domains. Future work could expand the exemplar pool with multilingual, cross-domain queries collected from paper-search platforms or directly contributed by researchers.

\section*{Ethical Considerations}
\label{sec:ethics}
\pname is constructed from publicly available scientific papers, bibliographic metadata, and \openreview discussions. The benchmark is
not designed to collect sensitive personal information. Our goal is to benefit the scientific research community.
Our human validators are instructed to avoid using  content from sources that prohibit copying or redistribution at the human cross-check stage for \pname dataset.
Consequently, most documents are derived from sources that are sources that permit research use or redistribution under their licenses, such as \openreview, Arxiv, ACL Anthology papers. The sampled human-written queries we leverage for query generation also carry permissive public licenses, including PaSa (CC-BY-4.0) and LitSearch (CC-BY-4.0).

\section*{Reproducibility}
Our data generation process is described in \Cref{sec:dataset}, with detailed information and prompt templates provided in Appendix \ref{appendix:maple_generation}. Moreover, experimental evaluation settings are discussed in \Cref{sec:experiments}, combined with details of retriever baselines in Appendix \ref{appendix:retriever_detail}. To facilitate the reproduction of our experiments, the data are available at \datasetlink, and the evaluation code is provided at \codelink.

\bibliographystyle{abbrvnat}
\bibliography{references}

\appendix

\section{The Use of Large Language Models}
In preparing this manuscript, large language models (LLMs) were utilized only for English grammar error detection and polishing. All substantive content and analyses were developed independently by the authors first. For dataset construction, GPT-5~\cite{openai2025gpt5} is used for query generation stage, moreover Deepseek~\cite{DeepSeekAI2026DeepSeekV4TH} is implemented in negative paper mining stage and the LLM-as-a-judge diagnostic stage.

\section{Analysis on human-written full-paper search query}
\label{appendix:analysis_human_query}
After collecting the human-written full-paper search queries (see \Cref{subsubsec:human_query}), to better understand the distribution of real scientific search intents, we analyze human-written paper-search queries from existing sources. We observe that real paper-search behavior varies along two dimensions:  aspect and granularity. In terms of aspect, queries may require different parts of a paper to judge relevance. For example, as in \Cref{tab:section_aspect_label_criteria}, query asking for papers that build dense retrievers with mixture-of-experts architectures targets method design, while a query asking for papers showing that smaller pre-training datasets can outperform larger datasets targets experimental findings.  In terms of granularity, queries range from broad topic-level discovery to fine-grained full-paper information needs. Broad topic-level queries for example "Give me papers about LLM quantized pretraining." are realistic and common, but they are closer to conventional ad-hoc scientific paper search, where short textual queries are matched against candidate papers represented primarily by titles, abstracts\cite{singh-etal-2023-scirepeval}. Also, they can admit many valid relevant papers, making sparse relevance labels vulnerable to false negatives under standard retrieval evaluation\cite{Upadhyay2024LLMsCP}.  In contrast, fine-grained queries with detailed conditions often require full-text evidence to judge relevance\cite{ajith-etal-2024-litsearch}. We therefore focus on fine-grained full-paper queries as the main scope of our benchmark. After careful selection, for motivation, method, experiment group, each include 11, 33, 40 human-written queries.

\section{\pname Generation Details}
\label{appendix:maple_generation}
\paragraph{Query Generation.}
We first use GPT-5.4 to summarize filtered \openreview comments into  aspect-specific bullet points using the prompt in \Cref{prompt:summarize_bullets} (multimodal version prompt in \Cref{prompt:summarize_bullets_multimodal}). We then generate query seeds, with prompt in \Cref{prompt:generate_seed} to select the most potential bullets in search needs and abstract it into natural language style. For retrieval-based in-context learning, we encode each seed with BGE-M3~\cite{chen-etal-2024-m3}, retrieve the top-$5$ human-written queries from the corresponding aspect group, and insert them as demonstrations into the query-generation prompt (\Cref{prompt:generate_ir}) to generate queries. Finally, using \Cref{prompt:fullpaper_retrieval_evaluation}, we filter queries that either exhibit excessive overlap with the target paper's title and abstract, suggesting target-specific copying, or can be resolved from this limited content without consulting the full paper.

\paragraph{Hard-Negative Mining.}
For each retained query, we use an LLM to extract search keywords with the prompt in \Cref{prompt:extract_keywords}. We submit these keywords to the Semantic Scholar bulk search API and retrieve the top-$100$ candidate papers. A reasoning LLM then reranks these candidates using \Cref{prompt:rerank}, after which we retain the top-$30$ candidates for full-text validation. For each candidate, we first use Docling~\cite{Auer2024DoclingTR} to extract the full-text content and then use 1M-token-context LLM Deepseek-V4-Pro to examine the complete paper with \Cref{prompt:judge_negative}, determining whether it satisfies the query. We only retain the true negative papers.



\section{Human Cross-Check Details}
\label{appendix:human_cross_check}
We conduct two complementary human evaluations to assess the quality of the automatically constructed benchmark: a pairwise query-realism evaluation and a query--paper relevance evaluation. Our annotator pool comprises five researchers with relevant domain expertise. Each annotator graduated from a university ranked among the top 300 in the 2026 QS World University Rankings and has published at least one machine learning conference paper.

\begin{table*}[t]
  \centering
  \small
  \begin{tabular}{p{0.18\textwidth}p{0.24\textwidth}p{0.48\textwidth}}
\toprule
\textcolor{black}{\textbf{Query Group}} & \textcolor{black}{\textbf{Original source}} & \textcolor{black}{\textbf{Query origin}} \\
\midrule
\pname (ours) & \methodname pipeline & Generated from \openreview comments and guided by real human queries. \\
Human Written & LitSearch-Human; \newline PaSa-RealScholarQuery & Human-written queries selected from the LitSearch and PaSa datasets. \\
LLM-generated + Human Editing & LitSearch-inline subset & GPT-4-generated from inline-citation contexts and manually examined or edited by experts. \\
LLM-generated & IRPAPERS benchmark & Claude Sonnet 4.5-generated page-level questions. \\
\bottomrule
\end{tabular}
  \caption{Query sources used in the blinded pairwise realism evaluation.}
  \label{tab:query_realism_source_labels}
\end{table*}

\subsection{Query Realism Evaluation}
We compare four query sources, summarized in \Cref{tab:query_realism_source_labels}: \pname queries, human-written queries, LLM-generated queries subsequently reviewed or edited by humans, and directly LLM-generated queries. We sample 150 \pname queries and construct three primary evaluation settings of 50 pairs each: \pname versus human-written, \pname versus human-edited LLM, and \pname versus directly LLM-generated queries. We additionally construct two calibration settings of 50 pairs each, comparing human-written queries with the two LLM-based sources. The evaluation therefore contains 250 query pairs across five comparison settings.

For each pair, the interface presents two queries as Query A and Query B without revealing their sources. Two annotators independently select one of three outcomes: Query A is more realistic, the two queries are equally realistic, or Query B is more realistic. Realism is defined as the likelihood that a researcher would naturally enter the query into a scientific paper-search system. A third annotator adjudicates disagreements between the initial judgments. Cohen's $\kappa$ is computed from the two initial independent annotations, whereas the preference percentages in \Cref{tab:query_realism_main} are calculated from the final adjudicated labels.

\subsection{Query--Paper Relevance Evaluation}
We sample 150 generated query--paper pairs to assess whether the multi-stage generation pipeline preserves the intended relevance relation. For each pair, evaluators inspect the generated query together with the full text of its target paper. A pair is labeled relevant when the paper contains sufficient evidence to satisfy the query's information need; otherwise, it is labeled irrelevant. This full-text protocol is necessary because many queries depend on detailed methods, experimental findings, figures, or tables that are not available in the title and abstract. Overall, 96.7\% of the sampled pairs are judged relevant, indicating limited semantic drift during query generation.

\section{LLM-as-a-Judge Query Style Diagnostics}
\label{app:llm_judge}
We use a rubric-based LLM judge to compare the stylistic properties of \methodname queries with human-written and directly LLM-generated queries. The judge evaluates three dimensions: (1) \textit{specificity calibration}, scored from 1 (overly broad) to 5 (overly detailed),using rubric prompt in \Cref{prompt:llm_judge_rubric_1_specificity}; (2) \textit{lexical naturalism}, scored from 1 (overly casual or fragmentary) to 5 (overly formal or document-like), using rubric prompt in \Cref{prompt:llm_judge_rubric_2_naturalism}; and (3) \textit{semantic constraint count}, defined as the number of distinct  conditions explicitly expressed in a query, using prompt in \Cref{prompt:llm_judge_rubric_3_constraint}. The complete judging instructions and output format are provided in \Cref{prompt:llm_judge}.

\section{Retriever Details}
\label{appendix:retriever_detail}
We list the model checkpoints corresponding to the dense retrievers used in our experiments in \Cref{tab:retriever_checkpoints}. For instruction-aware query encoding, we use the unified instruction “Represent this paper search query for retrieval of relevant scientific papers” for Qwen3-Embed-8B, Instructor-XL, GritLM-7B, Qwen3-VL-Embed-8B and Ops-MM-Embed-7B when encoding queries. For text-only retrieval, we evaluate BM25, Qwen3-Embed-8B, Instructor-XL, and GritLM-7B. For screenshot-based multimodal retrieval, we evaluate Qwen3-VL-Embed-8B and Ops-MM-Embed-7B.

\begin{table}[t]
  \centering
  \scriptsize
  \resizebox{\linewidth}{!}{\begin{tabular}{p{0.72\linewidth}r}
\toprule
    \textcolor{black}{\textbf{Primary Area}} & \textcolor{black}{\textbf{\%}} \\ \midrule
    Foundation or frontier models, including LLMs & 29.52 \\
    Generative models & 12.38 \\
    Datasets and benchmarks & 9.05 \\
    Reinforcement learning & 7.62 \\
    Applications to computer vision, audio, language, and other modalities & 6.67 \\
    Alignment, fairness, safety, privacy, and societal considerations & 4.29 \\
    Interpretability and explainable AI & 3.81 \\
    Other topics in machine learning & 3.81 \\
    Learning theory & 3.81 \\
    Unsupervised, self-supervised, semi-supervised, and supervised representation learning & 3.33 \\
    Others & 15.71 \\
    \bottomrule
\end{tabular}}
  \caption{Primary-area distribution of target ICLR 2026 papers used for query generation in \pname.}
  \label{tab:target_paper_primary_area}
\end{table}

\begin{table}[H]
    \centering
    \scriptsize
    \begin{tabular}{p{0.34\linewidth}p{0.58\linewidth}}
        \toprule
        \textbf{Retriever} & \textbf{HuggingFace Checkpoint / Model} \\
        \midrule
        SPECTER2 & \path{allenai/specter2_base} \\
        SciNCL & \path{malteos/scincl} \\
        Instructor-XL & \path{hkunlp/instructor-xl} \\
        BGE-M3 & \path{BAAI/bge-m3} \\
        Qwen3-Embed-8B & \path{Qwen/Qwen3-Embedding-8B} \\
        GritLM-7B & \path{GritLM/GritLM-7B} \\
        Qwen3-VL-Embed-8B & \path{Qwen/Qwen3-VL-Embedding-8B} \\
        Ops-MM-Embed-v1-7B & \path{OpenSearch-AI/Ops-MM-Embed-7B} \\
        \bottomrule
    \end{tabular}
    \caption{Retriever implementations and HuggingFace checkpoints used in our experiments.}
    \label{tab:retriever_checkpoints}
\end{table}

\begin{table*}[t]
    \centering
    \small
    \setlength{\tabcolsep}{3pt}
    \begin{tabular}{p{0.12\textwidth}p{0.07\textwidth}p{0.18\textwidth}p{0.19\textwidth}p{0.35\textwidth}}
        \toprule
        Label & \#Query & Core question & Related paper section & Query examples  \\
        \midrule
        Motivation & 48 & Why does this problem exist or matter? & Introduction, Related Work, Problem Statement & Papers that apply RLHF to address the hallucination problem in image and video description. \\
        \addlinespace
        Method & 105 & How does the approach work? & Method, Model, Architecture, Training, Pipeline & Are there any papers that build dense retrievers with mixture-of-experts architecture where each expert is responsible for different types of queries?\newline Give me papers about how to rank search results by the use of LLM. \\
        \addlinespace
        Experiment/result & 47 & What was tested and what did the results show? & Experiments, Results, Analysis, Ablations, Findings & Give me papers which show that using a smaller dataset in large language model pre-training can result in better models than using bigger datasets.\newline Provide papers demonstrating that the self-correction of LLMs does not enhance their performance. \\
        \bottomrule
    \end{tabular}
    \caption{Labeling criteria for the three coarse aspect groups in the human-written full-paper query pool. Each group indicates the type and sections of paper content that a retriever would need to inspect to determine relevance.}
    \label{tab:section_aspect_label_criteria}
\end{table*}

\begin{table*}[t]
  \centering
  \scriptsize
  \begin{tabular}{p{0.06\textwidth}p{0.09\textwidth}p{0.31\textwidth}p{0.23\textwidth}p{0.21\textwidth}}
    \toprule
    \textbf{Subset} & \textbf{Aspect Group} & \textbf{\openreview Comments} & \textbf{Summarized Bullet Points} & \textbf{Generated Query} \\
    \midrule
    \multirow{3}{*}{\shortstack{Text-\\referenced\\main set}} & Motivation & Reviews/Review 1/review: ``... \textcolor{BrickRed}{linear compute complexity and constant memory usage}, in contrast to standard LongCoT, which incurs \textcolor{BrickRed}{quadratic computational costs} ...'' \newline Reviews/Review 2/review: ``... the paper deals with efficiency in reasoning language models, particularly the \textcolor{BrickRed}{compute cost of long chain-of-thought} ... achieves more favorable compute scaling and \textcolor{BrickRed}{constant memory} ...'' & The paper addresses the inefficiency of standard \textcolor{BrickRed}{long chain-of-thought reasoning}, whose full-context self-attention causes \textcolor{BrickRed}{compute to grow quadratically} with reasoning length and makes \textcolor{BrickRed}{long-horizon inference and RL training expensive or impractical}. & Is there any paper on addressing the \textcolor{BrickRed}{compute and memory cost of long chain-of-thought reasoning} in language models? \\
    \cmidrule(l){2-5}
    & Method & Reviews/Review 3/review: ``... the policy is updated using \textcolor{BrickRed}{preference-based feedback} which takes the form of a \textcolor{BrickRed}{binary score between pairs of presented trajectories} ...'' \newline Comments/Comment 14/comment: ``... BRIDGE [is] the first rigorous theoretical framework for ... \textcolor{BrickRed}{offline imitation learning followed by online preference-based fine-tuning} ... we added new ablation studies ... \textcolor{BrickRed}{BRIDGE degrades gracefully with noisy feedback} and maintains lower regret than the baseline ...'' & The online phase uses \textcolor{BrickRed}{binary preference feedback} between presented \textcolor{BrickRed}{trajectories} to guide policy improvement. & Are there papers that learn control policies from \textcolor{BrickRed}{pairwise human preferences} over \textcolor{BrickRed}{trajectories}? \\
    \cmidrule(l){2-5}
    & Experiment/result & Reviews/Review 3/review: ``... \textcolor{BrickRed}{PPO-based RLHF} often narrows the model's output distribution ... the authors introduce Support Retention Ratio (SRR) ... \textcolor{BrickRed}{CaPPO} lifts \textcolor{BrickRed}{win rate by ~2--4 points} over \textcolor{BrickRed}{PPO} and raises \textcolor{BrickRed}{SRR by ~0.20--0.30}, while improving Distinct-2 and lowering Self-BLEU ...'' & Relative to \textcolor{BrickRed}{PPO, CaPPO} improves \textcolor{BrickRed}{win rate by roughly 2--4 points} while also increasing \textcolor{BrickRed}{SRR by about 0.20--0.30}. & Could you list research showing that \textcolor{BrickRed}{reinforcement learning based fine-tuning} can improve preference \textcolor{BrickRed}{win rate} while preserving more varied generations than standard \textcolor{BrickRed}{PPO} for language models? \\
    \midrule
    \multirow{3}{*}{\shortstack{Multimodal-\\referenced \\subset}} & Motivation & Reviews/Review 1/pros: ``... The paper convincingly motivates the two main issues (\textcolor{BrickRed}{search space truncation and error amplification}) with illustrative examples (\textbf{\textcolor{blue}{Figure 1}}). ...'' & \textbf{\textcolor{blue}{Figure 1}} is used to motivate PARoG by illustrating the two central failure modes in KG-augmented LLM reasoning: \textcolor{BrickRed}{search space truncation bias and entity error amplification}. & Which paper studies \textcolor{BrickRed}{common failure modes} of language models for knowledge graph question answering, especially on \textcolor{BrickRed}{multi-hop or logic-heavy questions}? \\
    \cmidrule(l){2-5}
    & Method & Reviews/Review 4/cons: ``... the \textcolor{BrickRed}{RSA metric} (\textbf{\textcolor{blue}{Eq. 1}}) ... is a \textcolor{BrickRed}{simple, linear metric} ... This use of a simple linear metric to solve a \textcolor{BrickRed}{complex non-linear problem} weakens the convincingness of RSA ...'' 
     \newline Comments/Comment 11/comment: ``... Appendix F validates the efficiency of the RSA metric, showing in \textbf{\textcolor{blue}{Table 17}} and \textbf{\textcolor{blue}{Figure 9}} that it captures the \textcolor{BrickRed}{same geometric structures as computationally expensive non-linear metrics} (k-NN, Silhouette) but with a \textcolor{BrickRed}{49x speedup}. ...'' & \textbf{\textcolor{blue}{Equation 1}} defines RSA as a centroid- and variance-based \textcolor{BrickRed}{linear separability metric}, which reviewers argue may be too simplistic to faithfully identify experts in the \textcolor{BrickRed}{complex manifold structures} the paper claims to handle. & Are there methods for vision anomaly detection that test whether \textcolor{BrickRed}{simple linear separability measures} can reliably identify useful internal features in pretrained multimodal models? \\
    \cmidrule(l){2-5}
    & Experiment/result & Reviews/Review 4/pros: ``... The performance gains, especially for \textcolor{BrickRed}{2-bit quantization, are dramatic} (\textbf{\textcolor{blue}{Table 2}}). The method \textcolor{BrickRed}{successfully retains performance} where \textcolor{BrickRed}{SOTA PTQ methods} (GPTQ, AWQ) \textcolor{BrickRed}{fail completely}. ...'' 
    & \textbf{\textcolor{blue}{Table 2}} is described as showing \textcolor{BrickRed}{dramatic 2-bit empirical gains over PTQ baselines} such as GPTQ and AWQ... 
    & Could you list research showing that \textcolor{BrickRed}{2-bit quantization} can preserve reasoning accuracy \textcolor{BrickRed}{better than common post-training quantization methods} for language models? \\
    \bottomrule
\end{tabular}
  \caption{Query examples comparing the text-referenced main set and the multimodal-referenced subset. The examples show the generated query, summarized bullet-point content grounded in  paper evidence, and the supporting \openreview comments.}
  \label{tab:query_generation_examples}
\end{table*}

\begin{table*}[t]
    \centering
    \scriptsize
    {\begin{tabular}{llllll}
    \toprule
    \textbf{Representation} & \textbf{Motivation R@20} & \textbf{Method R@20} & \textbf{Experiment/Result R@20} & \textbf{AllAspect@20} & \textbf{AspectCoverage@20} \\
    \midrule
    Abstract only & 77.68 & 64.21 & 51.21 & 11.90 & 60.27 \\
    \midrule
    Full text & 72.77 & 65.29 & 53.32 & 14.76 & 61.39 \\
    \quad -- chunking (max pooling) & 77.23 (+4.46) & 67.03 (+1.74) & 58.80 (+5.48) & 14.29 (-0.48) & 64.79 (+3.40) \\
    \quad -- chunking (top-3 avg pooling) & \underline{82.14} (+9.38) & \underline{73.64} (+8.35) & \underline{66.70} (+13.38) & \underline{24.29} (+9.52) & \textbf{71.84} (+10.45) \\
    \midrule
    Full text + OCR & 73.21 & 64.75 & 54.27 & 15.71 & 61.76 \\
    \quad -- chunking (max pooling) & 78.12 (+4.91) & 66.59 (+1.84) & 57.74 (+3.48) & 14.76 (-0.95) & 64.31 (+2.55) \\
    \quad -- chunking (top-3 avg pooling) & \textbf{83.04} (+9.82) & \textbf{74.08} (+9.33) & 65.12 (+10.85) & 22.38 (+6.67) & \underline{71.33} (+9.56) \\
    \midrule
    Full text + interleaved images & 49.11 & 42.84 & 37.72 & 6.67 & 41.66 \\
    \quad -- chunking (max pooling) & 75.89 (+26.79) & 65.29 (+22.45) & 56.59 (+18.86) & 13.81 (+7.14) & 62.82 (+21.17) \\
    \quad -- chunking (top-3 avg pooling) & 81.70 (+32.59) & 71.80 (+28.96) & 64.17 (+26.45) & 23.33 (+16.67) & 69.81 (+28.15) \\
    \midrule
    PDF screenshot & 59.82 & 61.50 & 52.69 & 11.43 & 57.27 \\
    \quad -- chunking (max pooling) & 71.43 (+11.61) & 64.32 (+2.82) & 58.38 (+5.69) & 16.19 (+4.76) & 62.66 (+5.39) \\
    \quad -- chunking (top-3 avg pooling) & 77.23 (+17.41) & 73.54 (+12.04) & \textbf{66.91} (+14.23) & \textbf{24.76} (+13.33) & 71.18 (+13.92) \\
    \bottomrule
\end{tabular}}
    \caption{Retrieval performance of different paper representations. Query-level Recall@20 is reported for motivation, method, and experiment/result queries, and paper-level metrics are reported with AllAspect@20 and AspectCoverage@20. Scores are percentages; values in parentheses show the difference from the corresponding non-chunked representation.}
    \label{tab:representation_recall}
\end{table*}

\begin{figure*}[p]
    \centering
    \begin{subfigure}{0.95\textwidth}
        \centering
        \includegraphics[width=\linewidth]{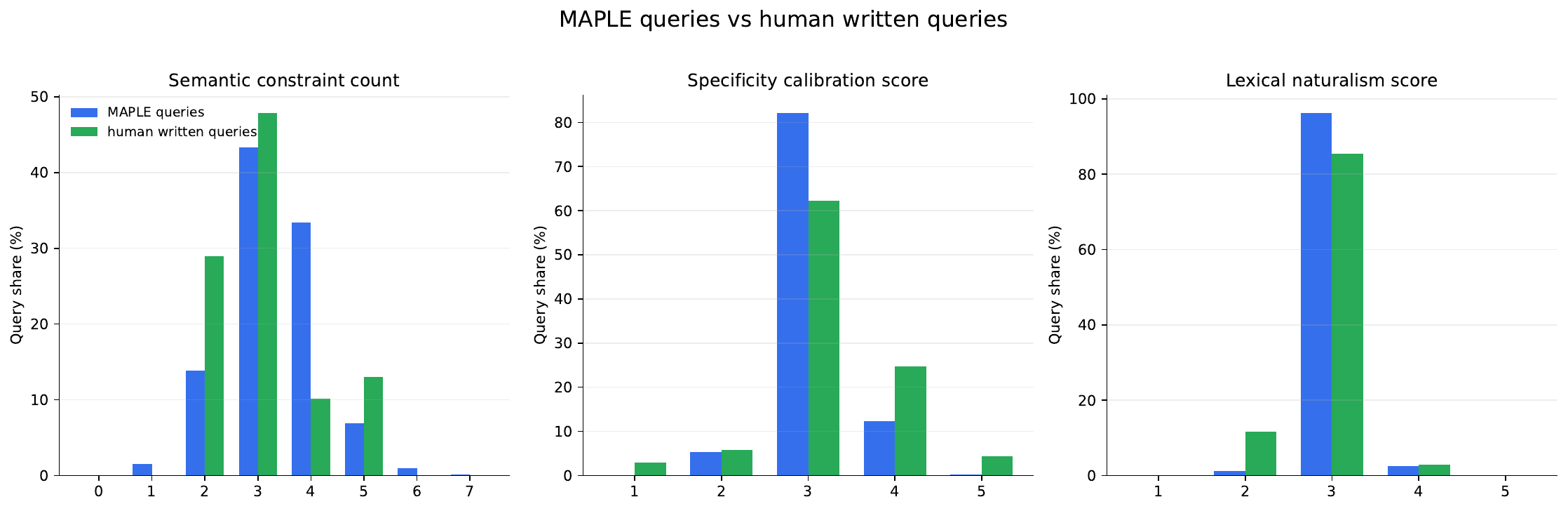}
        \caption{\methodname versus human-written queries.}
        \label{fig:maple_human_judge}
    \end{subfigure}

    \vspace{0.8em}

    \begin{subfigure}{0.95\textwidth}
        \centering
        \includegraphics[width=\linewidth]{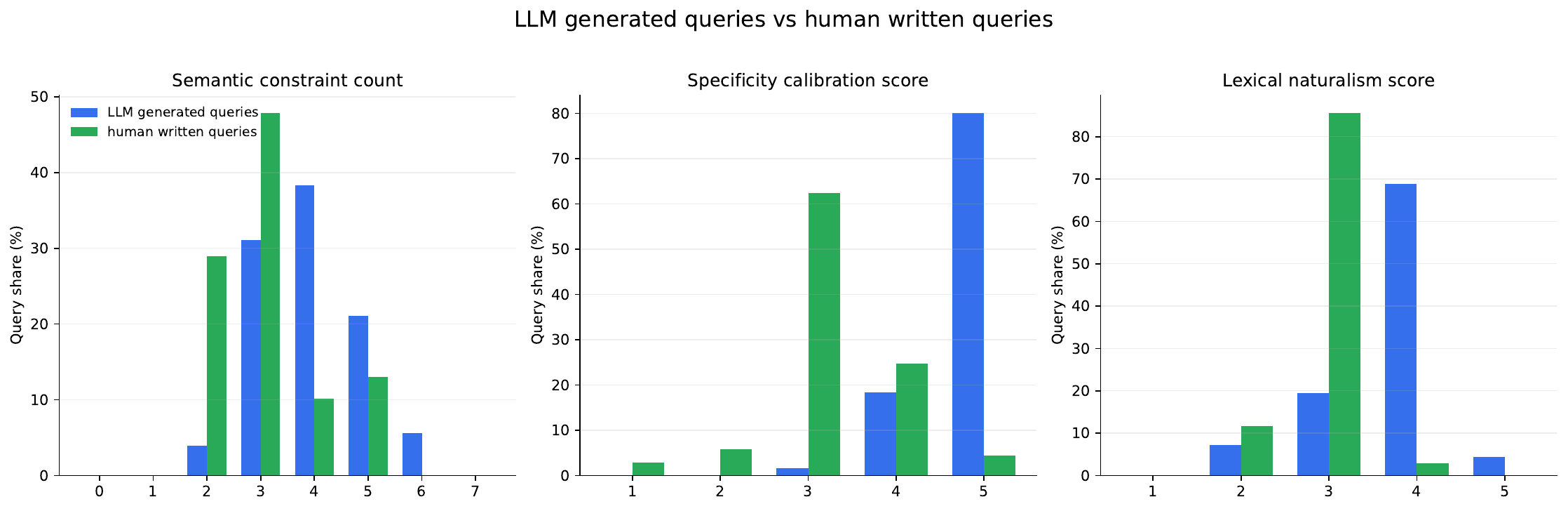}
        \caption{Directly LLM-generated versus human-written queries.}
        \label{fig:llm_human_judge}
    \end{subfigure}
    \caption{Distributions of LLM-judge-derived scores for specificity calibration, lexical naturalism, and semantic constraint count. \pname queries exhibit greater distributional overlap with human-written queries than directly LLM-generated queries, which tend toward greater specificity, formality, and constraint density.}
    \label{fig:llm_judge_comparison}
    \Description{Two vertically arranged groups of bar charts comparing query-style score distributions. The first compares MAPLE-Synth with human-written queries, and the second compares directly LLM-generated queries with human-written queries.}
\end{figure*}



\begin{table*}[t]
\centering
\begin{minipage}{0.98\textwidth}
\noindent\textcolor{black}{\rule{\linewidth}{1pt}}
\begin{lstlisting}[basicstyle=\ttfamily\scriptsize,breaklines=true,columns=fullflexible]
# Literature-Review Style Summarization

You are summarizing OpenReview content for one scientific paper.

Write clean literature-review notes about the paper itself, not a narration of the review conversation.

## Views
Organize the output into exactly these views:
- `motivation`
- `method`
- `experiment/result`

Use the views as follows:
- `motivation`: targets WHY this research problem exists and WHY this paper's approach is necessary, such as the research problem, need, gap, goal, hypothesis, or reason the work matters
- `method`: targets HOW the proposed approach works mechanistically, such as the proposed approach, model, algorithm, system, procedures, pipelines, dataset construction process, or implementation design, training/inference steps,
- `experiment/result`:  targets WHAT was empirically tested and what the results showed and analyzed, such as evaluation setup, benchmarks, datasets used for testing, metrics, baselines, ablations, empirical findings, measured performance, comparisons, analytical observations, and observed behavior, limitations.

Each view should contain:
- `summary`: a short summary of that view
- `bullet_points`: concise atomic claims for that view

Bullet count limits:
- `motivation`: generate at most 3 bullet points, select those most representative and distinctive of the paper's motivation
- `method`: generate at most 10 bullet points, select those most representative and distinctive of the paper's method, but do not include every single detail of the method, only the most important ones that are likely to be useful for retrieval
- `experiment/result`: generate at most 10 bullet points, select those most representative and distinctive of the paper's experiments and results, but do not include every single detail of the experiments, only the most important ones that are likely to be useful for retrieval

## Principles
- Merge duplicate reviewer/rebuttal/comment points into one clear claim.
- Put each claim in the single best-fit view (motivation, method, or experiment/result).
- Exclude procedural comments and thanks.
- If a concern is clarified by rebuttal/comments, write the final paper-level interpretation.
- If support is weak or disputed, use calibrated wording such as "The paper argues..." or "Evidence remains limited for..."
- Each OpenReview field is prefixed with a source path like `[Reviews/Review 1/pros]`.
- For every bullet point, include `source_refs` with the exact source paths that support the claim.

## Output Format
Return valid JSON only:

```
{{
  "motivation": {{
    "summary": "short summary for this view",
    "bullet_points": [
      {{
        "text": "one atomic literature-review claim",
        "source_refs": ["Reviews/Review 1/pros"]
      }}
    ]
  }},
  "method": {{
    "summary": "short summary for this view",
    "bullet_points": []
  }},
  "experiment/result": {{
    "summary": "short summary for this view",
    "bullet_points": []
  }}
}}
```

## Paper Information
Title: {paper_title}
Abstract: {paper_abstract}

## Full OpenReview Content
{full_openreview_content}

## Output (JSON)

\end{lstlisting}
\noindent\textcolor{black}{\rule{\linewidth}{1pt}}
\end{minipage}
\caption{Prompt for Summarizing OpenReview content for one scientific paper in a literature-review style.}
\label{prompt:summarize_bullets}
\end{table*}

\begin{table*}[t]
\centering
\begin{minipage}{0.98\textwidth}
\noindent\textcolor{black}{\rule{\linewidth}{1pt}}
\begin{lstlisting}[basicstyle=\ttfamily\scriptsize,breaklines=true,columns=fullflexible]

# Multimodal Evidence Bullet Generation

You are summarizing OpenReview discussion about concrete multimodal evidence items in a scientific paper.

Only use the provided snippets. Every evidence group contains snippets that explicitly mention the same figure/table/equation/algorithm item.
Do not use or infer from review/comment text outside the provided snippets.
Appendix references can count as multimodal evidence if the discussion is substantive. If an evidence group is not substantive, explicitly skip it.

For every evidence group in the input, make exactly one decision:
- generate exactly one literature-review-style bullet if the group contains substantive discussion, or
- add the group to `skipped_evidence_groups` if the snippets mention the evidence item but do not support a meaningful paper-level bullet.

Do not silently omit any evidence group.

## View labels
- `motivation`: why the problem matters, why the approach is needed, background, gaps, goals, or limitations motivating the work
- `method`: how the method/model/algorithm/dataset/system works
- `experiment/result`: evaluation setup, benchmarks, metrics, ablations, empirical findings, comparisons, trends, or limitations in results

## Dependency labels
- `incidental`: the evidence is mentioned but is not essential to the claim
- `supportive`: the evidence supports the claim, but the claim can still be understood from text
- `necessary`: the claim depends on this figure/table/equation/algorithm; removing it would weaken or invalidate the claim

Important: do not output bullets with `multimodal_dependency: "incidental"`.
If the best dependency label would be `incidental`, put that evidence group in `skipped_evidence_groups` instead.

## Rules
- Every evidence group must appear exactly once, either in `bullet_points` or in `skipped_evidence_groups`.
- Each output bullet must be grounded in one evidence group.
- Copy the evidence group's integer `evidence_group_id`.
- Copy the evidence group's exact label into `multimodal_ref`, for example `["Table 1"]`.
- Keep each bullet atomic: one paper-level claim, not a list of observations.
- Do not merge multiple evidence groups into one bullet.
- Include a concrete `multimodal_dependency_rationale` explaining what the evidence item contributes and whether text alone would support the claim.
- If a group has only a passing mention, procedural note, or unclear reference with no meaningful claim, skip it and explain why in `skip_rationale`.
- If the evidence is merely incidental to the claim, skip it and explain that the multimodal evidence is only incidental.

## Output
Return valid JSON only:

```json
{{
  "bullet_points": [
    {{
      "evidence_group_id": 1,
      "target_view": "experiment/result",
      "text": "one atomic paper-level claim grounded in one evidence group",
      "source_refs": ["Reviews/Review 1/weaknesses"],
      "multimodal_ref": ["Table 1"],
      "multimodal_dependency": "supportive",
      "multimodal_dependency_rationale": "Explain what the referenced evidence contributes and whether text alone would still support the claim."
    }}
  ],
  "skipped_evidence_groups": [
    {{
      "evidence_group_id": 2,
      "multimodal_ref": ["Figure 8"],
      "skip_rationale": "The snippets mention the figure but do not provide enough substantive paper-level discussion to form a reliable bullet."
    }}
  ]
}}
```

## Paper
Title: {paper_title}
Abstract: {paper_abstract}

## Evidence groups
Total groups: {evidence_group_count}

{evidence_groups}

## Output (JSON)


\end{lstlisting}
\noindent\textcolor{black}{\rule{\linewidth}{1pt}}
\end{minipage}
\caption{Prompt for summarizing multimodal evidence into atomic paper-level bullet points.}
\label{prompt:summarize_bullets_multimodal}
\end{table*}

\begin{table*}[t]
\centering
\begin{minipage}{0.98\textwidth}
\noindent\textcolor{black}{\rule{\linewidth}{1pt}}
\begin{lstlisting}[basicstyle=\ttfamily\scriptsize,breaklines=true,columns=fullflexible]

You are generating IR-style literature search queries from query-facing bullet seeds.

The input bullets have already been abstracted into `query_seed` text. Use those seeds as relevance anchors, not as final wording.

Use the retrieved golden queries as the main style guide. Generate all requested queries for this view in one JSON response.

## Goal
Write natural human search queries as if the user had not seen this paper.

Think in this direction:
scholar search need -> this source paper is relevant because the selected query_seed supports it.

## Style Rules
- Use retrieved golden queries as the main guide for wording, sentence shape, length, and specificity. Also use them to calibrate topic specificity -- golden queries describe research needs at a level any scholar in the field would recognize, using general method names, known benchmarks, and broad problem framings.
- Prefer words, phrases, and openings from the retrieved golden queries over words from the query_seed.
- Before writing each query, identify words in the seed that appear in the paper title or describe this paper's unique mechanism. Replace those with broader research terms an adjacent-field scholar would use.
- Target the middle level: broad enough to sound like a real search need, but specific enough that the selected query_seed is useful relevance evidence.
- Keep one plain full-text hook, such as a method family, dataset/setting, constraint, comparison, trade-off, ablation, or empirical behavior.
- Use everyday research wording. Keep standard field terms, but avoid paper-internal technical wording unless similar wording appears in the retrieved golden queries.
- Avoid reverse lookup queries: do not reconstruct this paper's exact mechanism, equation, assumption, implementation recipe, staged workflow, named component, parameter choice, baseline list, metric bundle, or table-specific detail.
- Avoid mini-paper titles, dense noun-phrase headers, exact numbers, paper-specific taxonomies, coined terms, step-by-step procedural chains, and exact figure/table/equation/algorithm labels.
- If the selected seed came from multimodal evidence, use that only to decide `is_multimodal` and the rationale. Do not put evidence-artifact words such as table, figure, equation, algorithm, formulation, approximation, or design into the query unless they appear in the retrieved golden queries.


## Generation Rules
- Generate exactly {num_queries} IR-style queries.
- Prefer different bullets; reuse a bullet only if there are too few strong seeds.
- If reusing a bullet, change the retrieval angle.
- Text-only bullets have `multimodal_ref: none`; set `is_multimodal=false` and `multimodal_rationale=null`.
- Multimodal bullets have a non-empty `multimodal_ref`; generate one query for each requested multimodal bullet, set `is_multimodal=true`, and use the bullet's multimodal rationale to explain why relevance depends on multimodal evidence.

Mode-specific rule:
{query_mode_instructions}

Retrieved IR-style exemplars:
{retrieved_examples}

Paper title:
{paper_title}

View label:
{view_label}

View definition:
{view_definition}

View-specific requirement:
{view_specific_requirement}

View summary:
{view_summary}

Query-facing bullet seeds:
{bullets}

Output valid JSON only:

{{
  "queries": [
    {{
      "query_text": "...",
      "is_multimodal": false,
      "multimodal_rationale": null,
      "related_bullet_indice": 1,
      "related_bullet_justification": "Briefly explain why the selected query_seed makes this paper relevant to the query."
    }},
    {{
      "query_text": "...",
      "is_multimodal": true,
      "multimodal_rationale": "Use the selected bullet's multimodal rationale to explain why relevance depends on multimodal evidence.",
      "related_bullet_indice": 2,
      "related_bullet_justification": "Briefly explain why the selected query_seed makes this paper relevant to the query."
    }}
  ]
}}





\end{lstlisting}
\noindent\textcolor{black}{\rule{\linewidth}{1pt}}
\end{minipage}
\caption{Prompt for generating IR-style literature search queries from query-facing bullets}
\label{prompt:generate_ir}
\end{table*}

\begin{table*}[t]
\centering
\begin{minipage}{0.98\textwidth}
\noindent\textcolor{black}{\rule{\linewidth}{1pt}}
\begin{lstlisting}[basicstyle=\ttfamily\scriptsize,breaklines=true,columns=fullflexible]
You are rewriting detailed paper-summary bullets into query-facing search seeds.

The seed is not a final query. It is a short search intent that will later be rewritten using golden query examples.

## Goal
For each bullet, write one query_seed that captures why this paper could be relevant to a scholar search in easy and natural language.

## Rules
- Simpler and more general than the bullet. Use everyday research wording; keep standard field terms only.
- Keep one concrete hook (method family, setting, constraint, comparison, trade-off, ablation, observed behavior).
- Sound like a search intent fragment, not a paper title.
- Strip anything that a scholar outside this paper's immediate subfield wouldn't recognize: paper-internal terminology, mechanism details, evaluation infrastructure words. If a word only makes sense after reading the paper, drop it.
###HARD RULE: Avoid overlap with the source paper abstract. If a seed would reuse abstract wording verbatim or as a close lexical variant, replace it with a natural paraphrased term or phrase instead , avoid explicit overlap


## Extra guidance
- Bullets about benchmarks / evaluation: reduce to the broad task category and one simple need. Don't carry structure descriptors as a bundle.
- Bullets about methods: describe the capability or broad method family, not how the paper achieves it internally.

Paper: {paper_title}
Abstract: {paper_abstract}
View: {view_label} -- {view_definition}
{view_specific_requirement}
Summary: {view_summary}

Bullets:
{bullets}

Output valid JSON only:
{{
  "seeds": [
    {{
      "bullet_index": 1,
      "query_seed": "plain abstracted search intent"
    }}
  ]
}}


\end{lstlisting}
\noindent\textcolor{black}{\rule{\linewidth}{1pt}}
\end{minipage}
\caption{Prompt for generating query-facing search seeds from detailed paper-summary bullets.}
\label{prompt:generate_seed}
\end{table*}

\begin{table*}[t]
\centering
\begin{minipage}{0.98\textwidth}
\noindent\textcolor{black}{\rule{\linewidth}{1pt}}
\begin{lstlisting}[basicstyle=\ttfamily\scriptsize,breaklines=true,columns=fullflexible]
Given the following retrieval query for a scientific paper, rewrite it into exactly 2 concise search phrases for paper retrieval.

Query: {{query}}

Requirements:
- Output exactly 2 phrases.
- Each phrase should be 2-4 words only.
- Prefer noun phrases that are easy for Semantic Scholar, Google Scholar, or arXiv to match.
- Keep complete technical terms intact when they are standard phrases.
- Keep only the 2 most representative paper family, task, method, or benchmark concepts.
- Drop extra explanations, conversational wording, and long natural-language constraints.
- Avoid quotes, punctuation, and full-sentence wording.
- The phrases should help find related but DISTINCT papers, not the exact same paper only.

Good style examples:
- "reasoning intensive retrieval"
- "multimodal retrieval benchmark"
- "parameter-efficient fine-tuning"
- "vision language action model"

Output a JSON array of strings only:
["phrase 1", "phrase 2"]
\end{lstlisting}
\noindent\textcolor{black}{\rule{\linewidth}{1pt}}
\end{minipage}
\caption{Prompt for extracting search keywords from a paper search query, using in the hard-negative mining stage.}
\label{prompt:extract_keywords}
\end{table*}

\begin{table*}[t]
\centering
\begin{minipage}{0.98\textwidth}
\noindent\textcolor{black}{\rule{\linewidth}{1pt}}
\begin{lstlisting}[basicstyle=\ttfamily\scriptsize,breaklines=true,columns=fullflexible]

You are reviewing one retrieved paper for a scientific paper retrieval query.

Your job is to classify the candidate paper into exactly one label:
- `positive`: the paper genuinely satisfies the query well enough to be treated as a positive match
- `hard_negative`: the paper is close enough to look plausible for the query, but it misses a key requirement and should be a challenging near-miss negative
- `ignored`: the paper is not close enough, or the evidence is too weak

Important rules:
- Base the decision primarily on the candidate paper evidence provided below. If the full PDF is attached, use that as the main evidence source.
- Use the title / venue / abstract only as supporting metadata.
- Do not mark a paper as `positive` unless the paper evidence gives solid support that the query is satisfied.
- Use `hard_negative` only when the paper is topically close and plausibly confusable.
- Keep the decision conservative: use `hard_negative` only when the candidate is a genuine near miss, not just vaguely related.
- Set `need_pro_review` to `true` when the evidence is ambiguous, borderline, internally conflicting, or too weak for a confident final judgment.
- Set `need_pro_review` to `false` when the evidence is clear enough that a stronger model re-check is unnecessary.

Original query:
{{query}}

Candidate metadata:
- Title: {{paper_title}}
- Authors: {{authors}}
- Year: {{year}}
- Venue: {{venue}}
- Abstract/snippet: {{abstract}}

Candidate paper evidence:
{{paper_evidence}}

Return valid JSON only:
{
  "label": "positive" | "hard_negative" | "ignored",
  "need_pro_review": true | false,
  "reason": "1-2 sentences explaining the decision with specific evidence from the paper."
}


\end{lstlisting}
\noindent\textcolor{black}{\rule{\linewidth}{1pt}}
\end{minipage}
\caption{Prompt for judging a candidate paper whether are truly negative to a scientific paper retrieval query by its full-paper evidence.}
\label{prompt:judge_negative}
\end{table*}

\begin{table*}[t]
\centering
\begin{minipage}{0.98\textwidth}
\noindent\textcolor{black}{\rule{\linewidth}{1pt}}
\begin{lstlisting}[basicstyle=\ttfamily\scriptsize,breaklines=true,columns=fullflexible]
# Retrieval Effectiveness Evaluation
You are evaluating generated retrieval queries for one scientific paper.
Given the paper abstract and a list of queries, judge whether each query needs full-paper content for retrieval.

## Principle
Judge whether a keyword-based retriever using only the abstract could find this paper for the query. This is a lexical and semantic overlap judgment, not an inference task.

Lexical Overlap Definition:
For each key retrieval term, count it as one lexical-overlap term if every content word in that term appears in the abstract verbatim or as a morphological variant (Synonyms, paraphrases, and conceptual equivalents are semantic overlap, not lexical). Ignore stopwords. Do not require the words to appear as one contiguous span.

When identifying key retrieval terms, keep them short and atomic. Prefer 2-4 word retrieval-critical phrases rather than long descriptive clauses. If a query contains a long clause, break it into smaller retrieval terms before applying the Lexical Overlap Definition.

Three levels, check lexical overlap first, then semantic:
- `HIGH-LEXICAL-OVERLAP`: more than 2 key retrieval terms satisfy the Lexical Overlap Definition. An abstract-only keyword retriever would succeed.
- `LOW-LEXICAL-OVERLAP`: less than or equally 2 key retrieval terms satisfy the Lexical Overlap Definition. A keyword retriever would miss the paper. Full-paper content is needed.
- `LOW-SEMANTIC-OVERLAP`: key terms are NOT in the abstract AND the angle or aspect is also NOT covered by the abstract. Both lexical and semantic overlap are low. Full-paper content is needed.

## Output Format
Return a JSON array with one object per query, in input order:

```json
[
  {
    "query": "original query text",
    "dimensions": {
      "abstract_relevance": "LOW-LEXICAL-OVERLAP"
    },
    "reasoning": "1) Key retrieval terms: [list them]. 2) Per-term check: for each term, does it satisfy the Lexical Overlap Definition? State which content words are matched in the abstract and count how many terms satisfy the rule. 3) Angle check: does the abstract discuss the same aspect? 4) Conclusion"
  }
]
```

Use only `HIGH-LEXICAL-OVERLAP`, `LOW-LEXICAL-OVERLAP`, or `LOW-SEMANTIC-OVERLAP` for `abstract_relevance`.
Return valid JSON only. Do not add explanations before or after the JSON.

## Abstract

{{abstract}}

## Queries

{{queries}}
\end{lstlisting}
\noindent\textcolor{black}{\rule{\linewidth}{1pt}}
\end{minipage}
\caption{Prompt for evaluating whether retrieval queries require full-paper content.}
\label{prompt:fullpaper_retrieval_evaluation}
\end{table*}

\begin{table*}[t]
\centering
\begin{minipage}{0.98\textwidth}
\noindent\textcolor{black}{\rule{\linewidth}{1pt}}
\begin{lstlisting}[basicstyle=\ttfamily\scriptsize,breaklines=true,columns=fullflexible]

You are evaluating the relevance of research papers to a search query.

Query: {query}

Below are {len(candidates)} candidate papers. For each, judge how relevant it is
to the query on a scale of 1-10 (10 = directly addresses the query's core question,
1 = unrelated topic). Base your judgment only on the title and abstract snippet provided.

Candidates:
[{i}] {title}
    Abstract: {abstract}

Output a JSON array of objects with 'index' and 'score' fields, sorted by score descending.
Only include the top {top_n} candidates.
Format: [{"index": 0, "score": 9}, ...]


\end{lstlisting}
\noindent\textcolor{black}{\rule{\linewidth}{1pt}}
\end{minipage}
\caption{Prompt for reranking retrieved paper candidates from semantic scholar search.}
\label{prompt:rerank}
\end{table*}

\begin{table*}[p]
\centering
\begin{minipage}{0.98\textwidth}
\noindent\textcolor{black}{\rule{\linewidth}{1pt}}
\begin{lstlisting}[basicstyle=\ttfamily\tiny,breaklines=true,columns=fullflexible]
You are an expert judge of academic paper-search queries.

You will score each query on three metrics: Specificity Calibration, Lexical Naturalism, and Semantic Constraint Count.
Read the metric definitions and scoring rubrics carefully before scoring.

Your Task
Given a query, score it on each metric independently.
- For Specificity Calibration: output `score` and `rationale`
- For Lexical Naturalism: output `score` and `rationale`
- For Semantic Constraint Count: output `count` and `rationale`

Judge only style and human-likeness.
Do NOT judge retrieval correctness, paper relevance, or whether the paper actually exists.
Focus only on:
1. Specificity Calibration
2. Lexical Naturalism
3. Semantic Constraint Count

Important:
- Both scored metrics use a bipolar 1-5 scale.
- For both scored metrics, score 3 is ideal.
- Score 1 and score 5 are opposite failure modes.
- Do not assume that a higher score is better.
- The examples are illustrative anchors, not templates to match literally.

Keep the three metrics distinct:
- Specificity Calibration = whether the amount of detail is calibrated like a human query
- Lexical Naturalism = whether the wording and register sound like a human query
- Semantic Constraint Count = how many distinct retrieval-narrowing constraints are explicitly present

A query can score:
- high on one metric and low on another
- differently across metrics for different reasons

{{Criteria of Metric1. Specificity Calibration}}
{{Criteria of Metric2. Lexical Naturalism}}
{{Criteria of Metric3. Semantic Constraint Count}}

Queries:
{{numbered_queries}}

Return JSON only in this exact format:
{
  "results": [
    {
      "query": "...",
      "metric_1_specificity_calibration": {
        "score": 3,
        "rationale": "1-2 sentences citing specific textual evidence from the query."
      },
      "metric_2_lexical_naturalism": {
        "score": 2,
        "rationale": "1-2 sentences citing specific textual evidence from the query."
      },
      "metric_3_semantic_constraint_count": {
        "count": 2,
        "rationale": "1-2 sentences explaining the distinct retrieval constraints counted in the query."
      }
    }
  ]
}
\end{lstlisting}
\noindent\textcolor{black}{\rule{\linewidth}{1pt}}
\end{minipage}
\caption{Overall instructions and output format for evaluating the style and human-likeness of academic paper-search queries.}
\label{prompt:llm_judge}
\end{table*}

\begin{table*}[p]
\centering
\begin{minipage}{0.98\textwidth}
\noindent\textcolor{black}{\rule{\linewidth}{1pt}}
\begin{lstlisting}[basicstyle=\ttfamily\tiny,breaklines=true,columns=fullflexible]
Metric 1. Specificity Calibration
Definition:
Is the level of detail calibrated like a human-written researcher query?

What to judge:
- whether the query gives enough detail to guide retrieval
- whether it avoids being too broad
- whether it avoids becoming so detailed that it looks like reconstruction of a known paper rather than genuine search

Interpret this as a centered scale:
1 = far too broad
2 = somewhat too broad
3 = well calibrated
4 = somewhat too specific
5 = pathologically over-specific

Score 1 - Severely Under-specified
Description:
The query is so broad that it cannot guide retrieval in a meaningful way. It names a topic or domain but gives no useful filtering criterion.
Markers:
- almost any paper in the field would count
- no operational narrowing condition
- reads like a topic label or keyword area
Example:
"Give me papers about large language models."

Score 2 - Under-specified
Description:
The query identifies a real research direction, but lacks enough detail to distinguish the papers the researcher actually wants from a large amount of related work.
Markers:
- one broad topic is present
- some filtering is implied, but still too weak
- method, task, or setting is mentioned, but not enough to focus results well
Example:
"Give me papers about how to rank search results by the use of LLM."

Score 3 - Well-calibrated
Description:
The query is specific enough to constrain retrieval meaningfully, but not so narrow that it presupposes the answer. It reflects genuine uncertainty: the researcher knows what kind of paper they want, but not which paper.
Markers:
- multiple meaningful constraints interact
- the query implies a category of papers, not just one known paper
- results would need evaluation, not mere lookup
Example:
"Are there any large-scale and open-source text simplification datasets dealing with long passages?"

Score 4 - Over-specified
Description:
The query contains enough constraints that it starts to look like external reconstruction of a particular paper or a tiny handful of papers, rather than open-ended search.
Markers:
- several highly specific constraints
- unlikely combination of details
- reads more like remembered paper content than a normal search
Example:
"Find the NLP paper that focuses on dialogue generation and introduces advancements in the augmentation of one-to-many or one-to-one dialogue data by conducting augmentation within the semantic space."

Score 5 - Pathologically Over-specified
Description:
The query encodes so many specific details, exclusions, or methodological requirements that it reads like a structured filter over already-known papers, not a real search under uncertainty.
Markers:
- multiple detailed technical requirements across several dimensions
- exclusions or filtering clauses that feel review-like
- reads like a paraphrased abstract, requirements spec, or literature-review entry
Example:
"I am looking for the paper that builds a multimodal foundation model with visual, audio, and audio-visual pretraining data, excludes survey papers, and evaluates long-form audiovisual understanding across multiple benchmarks."
\end{lstlisting}
\noindent\textcolor{black}{\rule{\linewidth}{1pt}}
\end{minipage}
\caption{Detailed scoring rubric for Metric 1: Specificity Calibration.}
\label{prompt:llm_judge_rubric_1_specificity}
\end{table*}

\begin{table*}[p]
\centering
\begin{minipage}{0.98\textwidth}
\noindent\textcolor{black}{\rule{\linewidth}{1pt}}
\begin{lstlisting}[basicstyle=\ttfamily\tiny,breaklines=true,columns=fullflexible]
Metric 2. Lexical Naturalism
Definition:
Is the vocabulary, phrasing, and register consistent with how researchers actually write queries?

What to judge:
- whether the query sounds like something a real researcher would type
- whether the syntax is natural and query-like
- whether the register is appropriately academic without becoming essay-like or LLM-polished

Interpret this as a centered scale:
1 = keyword dump / fragmented
2 = awkward or stilted
3 = natural researcher register
4 = over-formalized / essay-like
5 = synthetic fluent / LLM-polished

Score 1 - Keyword Dump / Machine-like Fragmentation
Description:
The query is just a string of terms with no real syntactic structure. It reads like extracted keywords or metadata, not natural language.
Markers:
- no real sentence structure
- no connective logic
- noun phrases only
- may resemble SEO tags or structured search fragments
Example:
"LLM agent reinforcement learning reward shaping training evaluation benchmark"

Score 2 - Stilted / Over-compressed Natural Language
Description:
The query uses natural language, but it is awkward, compressed, or slightly malformed in ways that make it feel clunky rather than fluent.
Markers:
- missing articles or prepositions
- awkward conversational filler
- grammatically off or compressed phrasing
- recognizably human, but not well-formed
Example:
"Do you know some papers about using reward shaping methods to train large language model agent."

Score 3 - Natural Researcher Register
Description:
The query reads like something a researcher would actually type or say. The vocabulary is appropriately technical without being performatively formal.
Markers:
- direct and purposive
- fluent but not polished
- technical terms used naturally
- feels like someone searching while doing work
Example:
"Are there any papers that build dense retrievers with mixture-of-experts architecture where each expert is responsible for different types of queries?"

Score 4 - Over-formalized / Essay-register
Description:
The query is grammatically correct and technically fluent, but written in a register that is too formal or composed for search behavior. It sounds like prose, not a query.
Markers:
- full sentence or multi-sentence structure
- polite instruction wording
- requirement-spec style
- reads like writing rather than searching
Example:
"I am looking for research papers on the construction of multimodal foundation models that support both visual and audio inputs. These models should be pre-trained on large-scale datasets, including visual, audio, and audio-visual data. Please exclude survey papers."

Score 5 - Synthetic Fluency / LLM-polished Prose
Description:
The query is too clean, complete, and balanced to feel like real human search behavior. It reads like an LLM completing a prompt to write an idealized search query.
Markers:
- perfectly balanced clause structure
- every dimension of the topic is explicitly accounted for
- unusually complete, polished, and symmetric
- lacks the roughness typical of real human search phrasing
Example:
"Please provide scholarly works demonstrating that smaller, carefully curated pre-training datasets can yield superior large language models relative to larger corpora."
\end{lstlisting}
\noindent\textcolor{black}{\rule{\linewidth}{1pt}}
\end{minipage}
\caption{Detailed scoring rubric for Metric 2: Lexical Naturalism.}
\label{prompt:llm_judge_rubric_2_naturalism}
\end{table*}

\begin{table*}[p]
\centering
\begin{minipage}{0.98\textwidth}
\noindent\textcolor{black}{\rule{\linewidth}{1pt}}
\begin{lstlisting}[basicstyle=\ttfamily\scriptsize,breaklines=true,columns=fullflexible]
Metric 3. Semantic Constraint Count
Definition:
Count how many distinct retrieval-narrowing semantic constraints are explicitly present in the query.

What counts as a constraint:
- a task requirement
- a method or architecture requirement
- a dataset, modality, language, domain, or setting restriction
- a temporal, scale, robustness, efficiency, or supervision condition
- a first/earliest/originality condition
- an exclusion or negation clause

Counting guidance:
- Count semantically distinct constraints, not surface phrases.
- Do not count generic filler like "paper", "research", or "study".
- Do not split one tightly bound idea into multiple counts unless the query clearly separates them.
- If a query has no meaningful narrowing condition beyond a broad topic, return `0`.
- Return a non-negative integer only.
\end{lstlisting}
\noindent\textcolor{black}{\rule{\linewidth}{1pt}}
\end{minipage}
\caption{Detailed counting rubric for Metric 3: Semantic Constraint Count.}
\label{prompt:llm_judge_rubric_3_constraint}
\end{table*}

\end{document}